\documentclass[twocolumn,showpacs,superscriptaddress,aps,prd,%
longbibliography,nofootinbib]{revtex4-1}

\newcommand{\dd}{\mathrm d}

\newcommand{\ii}{\mathrm i}

\newcommand{\calL}{\mathcal L}

\usepackage{epsf}
\usepackage{amsmath}
\usepackage{amsfonts}
\usepackage{amssymb}
\usepackage{bm}
\usepackage{bbm}
\usepackage{nicefrac}
\usepackage{color}
\usepackage{pifont}
\usepackage{graphicx}
\usepackage{enumerate}
\usepackage{dcolumn}
\usepackage{maybemath}
\usepackage{siunitx}
\usepackage{xcolor}

\usepackage{slashed}
\DeclareMathOperator{\Tr}{Tr}

\newcommand\Refr[1]     {Ref.~\cite{#1}}

\newcommand\TODO[1]		{}

\newcommand\del			{\delta}
\newcommand\tht			{\theta}

\newcommand\vp			{\vec{p}}
\newcommand\avp[1]		{|\vec{p}_{#1}|}

\begin{document}

\newcommand{\addrMST}{Department of Physics, 
Missouri University of Science and Technology, Rolla, Missouri 65409, USA}

\newcommand{\addrDEB}{University of Debrecen, P.O.Box 105, H-4010 Debrecen, Hungary}

\newcommand{\addrMTADE}{MTA--DE Particle Physics Research Group, 
P.O.~Box 51, H--4001 Debrecen, Hungary}

\newcommand{\addrMTA}{MTA Atomki, P.O.~Box 51, H--4001 Debrecen, Hungary}

\title{Neutrino Splitting for Lorentz--Violating Neutrinos: Detailed Analysis}

\author{G. Somogyi}
\email{email: gabor.somogyi@cern.ch}
\affiliation{\addrMTADE}

\author{I. N\'andori}
\affiliation{\addrMTADE}
\affiliation{\addrDEB}
\affiliation{\addrMTA}

\author{U. D. Jentschura}
\email{email: ulj@mst.edu}
\affiliation{\addrMTADE}
\affiliation{\addrMTA}
\affiliation{\addrMST}

\begin{abstract}
Lorentz-violating neutrino parameters have been severely constrained
on the basis of astrophysical considerations.
In the high-energy limit,
one generally assumes a superluminal dispersion relation
of an incoming neutrino of the  form $E \approx |\vec p| \, v$,
where $E$ is the energy, $\vec p$ is the momentum and 
$v = \sqrt{1 + \delta}>1$.
Lepton-pair creation due to a Cerenkov-radiation-like 
process ($\nu \to \nu + e^- + e^+$) becomes possible above a certain
energy threshold, and bounds on the Lorentz-violating
parameter $\delta$ can be derived. 
Here, we investigate a related process,
$\nu_i \to \nu_i + \nu_f + \overline\nu_f$,
where $\nu_i$ is an incoming neutrino mass eigenstate,
while $\nu_f$ is the final neutrino mass eigenstate,
with a superluminal velocity that is slightly 
slower than that of the initial state.
This process is kinematically allowed if the 
Lorentz-violating parameters 
at high energy differ for the 
different neutrino mass eigenstates.
Neutrino splitting is not subject to any significant
energy threshold condition and could yield 
quite a substantial contribution to decay and energy loss 
processes at high energy, even if the differential 
Lorentz violation among neutrino flavors is severely 
constrained by other experiments.
We also discuss the $SU(2)_L$-gauge 
invariance of the superluminal models and
briefly discuss the use of a generalized {\em vierbein}
formalism in the formulation of the Lorentz-violating
Dirac equation.
\end{abstract}

\pacs{95.85.Ry, 11.10.-z, 03.70.+k}

\maketitle


%
%
\section{Introduction}
\label{sec1}

\color{black}
A possible Lorentz violation in the neutrino sector
has been the subject of intense investigations in recent
years, with a rich texture of interesting models and 
corresponding scenarios having been explored in the 
literature (see Refs~\cite{CoKo1998,KoMe2009,DiKoMe2009,KoMe2012,%
DiKoMe2014,Di2014,Ta2014,StEtAl2015,Li2015}).
\color{black}
The apparent cutoff of the neutrino spectrum seen
by IceCube at a threshold
energy $E = E_{\rm th} 
\approx \SI{2}{PeV}$~\cite{AaEtAl2013,AaEtAl2014,Bo2015}
has given rise to 
interesting speculations, which include 
rather stringent limits on 
Lorentz-violating parameters~\cite{St2014,StSc2014,StEtAl2015}.

Due to their different masses, neutrino decays
between generations are in fact kinematically
allowed, but for neutrinos which fulfill a 
Lorentz-invariant dispersion relation, the 
decays are excessively long and exceed the
age of the Universe by orders of magnitude.
Namely, in addition to the small magnitude of the 
mass differences, the 
Glashow--Iliopoulos--Maiani (GIM) mechanism~\cite{PaWo1982,GlIlMa1970}
leads to cancellations between generations. For quarks, 
the GIM mechanism relies on the unitarity of the 
Cabibbo--Kobayashi--Maskawa matrix, whereas for neutrinos, 
one assumes unitarity of the Pontecorvo--Maki--Nakagawa--Sakata
(PMNS) matrix. However, under a (small) Lorentz violation,
additional decay channels exist which are not GIM--suppressed,
and lead to decay of a neutrino without changing the 
flavor or mass eigenstates. Among these,
two decay and energy loss mechanisms 
have been given special attention,
namely, VPE (vacuum pair emission), which corresponds to 
lepton-pair Cerenkov radiation
(LPCR, $\nu \to \nu + e^- + e^+$),
and neutrino splitting,
i.e., neutrino-pair Cerenkov radiation
(NPCR, $\nu_i \to \nu_i + \nu_f + \bar \nu_f$).
We will use both designations 
for each process interchangeably in the current article.

Concerning NPCR, it has been pointed out
at various places in the 
literature~\cite{BeLe2012,MaLiMa2013,StEtAl2015}
that this process should be
suppressed, because all neutrino flavors are known to 
propagate at approximately the same velocity.
The papers~\cite{BeLe2012,AAEtAl2013boone} cite bounds on the
relative difference of the velocities of 
neutrino mass eigenstates on the level of $10^{-19}$,
based on the argument that otherwise, the interference
pattern of the neutrino oscillations would be smeared.
The short-baseline experiment~\cite{AAEtAl2013boone} has the advantage
that the entire beamline is under laboratory control.
The paper~\cite{CoGl1999} gives stricter
bounds on the order of $10^{-22}$, also from short-baseline experiments.
According to Ref.~\cite{AaEtAl2010lorentz}, the IceCube collaboration
has obtained, for the $\mu$-$\tau$ neutrino sector,
even stricter bounds on the 
order of $10^{-27}$,
coming from the survival of atmospheric muon neutrinos
in the energy range $100 \, {\rm GeV}$ to $10\,{\rm TeV}$,
assuming maximal mixing for some of the neutrino
flavor eigenstates.

If all neutrino propagation velocities
are precisely equal to each other, then the NPCR
process is kinematically forbidden~\cite{StEtAl2015}
in the Lorentz-violating Standard Model Extension,
and it seems that theoretical studies have thus focused
on Planck-scale operators inducing the NPCR decay.
These typically entail formulas for the decay rate which scale
with the eighth and higher powers of the neutrino energy 
[see Eqs.~(19) and~(20) of Ref.~\cite{StEtAl2015}]. 
The corresponding $\delta$ parameter
are proportional to the inverse of the Planck mass for
dimension-five operators [see Eqs.~(6) and (10) of 
Ref.~\cite{StEtAl2015}, the
latter with $n=1$] and proportional to the inverse of the square
of the Planck mass for dimension-six operators [see Eqs.~(6) and
(10) of Ref.~\cite{StEtAl2015}, 
the latter with $n=2$, as well as Refs.~\cite{MaEtAl2010jcap,MaLiMa2013}].
On one hand, one can argue that a detailed
analysis of the NPCR process within the 
Lorentz-violating Standard Model Extension is thus superfluous.
On the other hand, the NPCR process is unique 
among the decay processes for superluminal neutrinos
in that the energy threshold is negligible~\cite{StEtAl2015}.
It could thus be worthwhile to 
augment several treatments, recorded in the 
literature~\cite{CaCo2011,CiEvBiZh2012,HuEtAl2012},
by a detailed calculation of the effect,
in the spirit of Ref.~\cite{BeLe2012}, which includes
an analysis of the model dependence.

In order to put things into 
perspective, let us recall that the
VPE (or LPCR) decay rate is proportional to~\cite{CoGl2011,BeLe2012}
\begin{equation}
\label{expectation}
\Gamma_{\nu \to \nu e^- e^+} 
\propto G_F^2 E^5 (\delta_\nu - \delta_e)^3,
\end{equation}
where $\delta_\nu = v_\nu^2 - 1 \approx 
2 \, (v_\nu - 1)$ is the Lorentz-violating 
parameter of the incoming neutrino
(of energy $E$), and 
$\delta_e$ is the corresponding parameter
for the electron-positron pair.
Note that the 
$v_\nu$ parameter here takes the 
role of a maximum achievable velocity,
{\color{black}
consistent with the models proposed in 
Refs.~\cite{CoGl2011,BeLe2012}.}
(Note that units with $\hbar = c = \epsilon_0 = 1$ are
used throughout this paper.)
The corresponding parameters
in Refs.~\cite{StSc2014} and~\cite{StEtAl2015}
are defined for the deviations 
$v_\nu -1$ and $v_e - 1$ and thus 
differ from ours by a factor two.
The energy loss rate is found to be~\cite{CoGl2011,BeLe2012}
\begin{equation}
\frac{\dd E}{\dd x} \propto 
G_F^2 E^6 (\delta_\nu - \delta_e)^3 \,.
\end{equation}
Canonically~\cite{StSc2014,StEtAl2015},
one then makes the additional assumption that
$\delta_e = 0$, and justifies this on account of the
known subluminal nature of the electron
at low energies.
Furthermore, as pointed out in Refs.~\cite{St2014,StSc2014},
one has a bound on the order
of $\delta_e \leq 1.04 \times 10^{-20}$ for the 
Lorentz-violating parameter in the 
electron-positron sector.
(The bound is actually given as $0.52 \times 10^{-21}$
in Refs.~\cite{St2014,StSc2014}, but we again 
{\color{black} recall} the 
additional factor two.)

If one assumes that $\delta_e > 0$ could also
be superluminal for electrons in the high-energy region, then
the flavor-independent $\delta_\nu$
for neutrinos could be as large as
$\delta_\nu \approx 2.0 \times 10^{-20}$,
and still be compatible with the
bound on neutrinos from astrophysics,
and with the bounds on the electron
(see p.~7 of Ref.~\cite{StEtAl2015}).

In Ref.~\cite{StEtAl2015}, 
the NPCR process in the formulation adopted here, 
was simply discarded
on the basis of the argument that it vanishes
when all neutrinos propagate at the
same speed. We still found it interesting
to carry out a more detailed analysis of the
NPCR, in order to map out a possible role
of this effect in the analysis of
astrophysical data.

\color{black}
Based on an obvious analogy with Eq.~\eqref{expectation},
one might expect a functional form
\begin{equation}
\Gamma_{\nu_i \to \nu_i \nu_f \bar \nu_f} \propto 
G_F^2 E^5 (\delta_i - \delta_f)^3 \,,
\end{equation}
for the NPCR-induced decay rate,
where $\delta_i$ and $\delta_f$ refer to the
Lorentz-violating parameters of the initial
and final states. 
\color{black}
However, we can anticipate here
that one also finds terms of the form
\begin{equation}
\Gamma_{\nu_i \to \nu_i \nu_f \bar \nu_f} \propto 
G_F^2 E^5 (\delta_i - \delta_f) \, (\delta_i + \delta_f)^2,
\end{equation}
which could play a much more prominent role in the
analysis of astrophysical data than previously
thought, because the NPCR process essentially
has negligible threshold. (This assumption is made
in Ref.~\cite{StEtAl2015} and here.)
For the energy loss rate, we also find terms of the form
\begin{equation}
\frac{\dd E}{\dd x} \propto G_F^2 E^6 (\delta_i - \delta_f) \, 
(\delta_i + \delta_f)^2 \,,
\end{equation}
which could also be relevant in the high-energy region.

\color{black}
Some remarks on the gauge structure 
of the models employed in Refs.~\cite{CoGl2011,BeLe2012}
are in order. Of course, {\em a priori}, one would 
like to preserve the $SU(2)_L \times U(1)_Y$ gauge 
structure as much as possible upon the introduction
of the Lorentz-breaking parameters.
In addition, an inspection of the Lagrangian of the 
Standard Model Extension (SME) [see Eqs.~(9) and (10) 
of Ref.~\cite{CoKo1998}] reveals that 
the SME is based on the assumption that 
the entire gauge structure of the Standard Model
is preserved upon the introduction of the Lorentz-violating 
terms. 

In terms of the formulation of the vacuum pair emission~\cite{CoGl2011,BeLe2012},
one faces a certain dilemma: On one hand, if the 
charged fermions and the neutrinos are grouped in an
$SU(2)_L$ doublet, {\em and} if the Lorentz-breaking parameters
are assigned uniformly over all generations, then both 
NPCR and LPCR decays are kinematically forbidden
in the Lorentz-violating Standard Model Extension.
On the other hand, there are good reasons (with both 
theoretical as well as experimental motivations) for the 
assumption that Lorentz violation, if it exists, 
should be confined to the neutrino sector, 
without affecting the charged fermions.

Hence, Cohen and Glashow, as well as Bezrukov and Lee~\cite{CoGl2011,BeLe2012}
(see also Sec.~VII C 2 of Ref.~\cite{KoMe2012}),
chose interaction Lagrangians which, as a closer
inspection reveals (see also Appendices~\ref{appa} and~\ref{appb},
Sec.~VII C 2 of Ref.~\cite{KoMe2012},
and Ref.~\cite{JeNaSo2019}), are {\em not} gauge invariant
with respect to the full $SU(2)_L \times U(1)_Y$ gauge group.
We here follow the same approach, and, in addition to the 
analysis of the NPCR decay, attempt to find certain
interpolating formulas connecting the models 
used in Refs.~\cite{CoGl2011,BeLe2012}.
Some of the models used by Bezrukov and Lee
[see the cryptic remark on
``gauge invariance'' near Eq. (4) of Ref.~\cite{BeLe2012}]
preserve the gauge structure of the electroweak 
interaction at least in part (see also Ref.~\cite{JeNaSo2019}).

The models used in Refs.~\cite{CoGl2011,BeLe2012} and 
here are based on two assumptions, namely,
{\em (i)} that Lorentz-violation is confined to the
neutrino sector, and that
{\em (ii)} the effective Fermi theory, possibly with some
modifications,  still holds for
the description of the decay and energy loss processes.
We here follow this approach,
which was also used in Sec.~VII C 2 of
Ref.~\cite{KoMe2012},
and accept a (perturbative, see Appendix~\ref{appb}) 
violation of $SU(2)_L$ gauge invariance
as a price for the attractive Lorentz-violating 
kinematic scenario under investigation.
\color{black}

This paper is organized as follows.
In Sec.~\ref{sec2}, we discuss the framework
of the calculation, partially outlined above,
in greater detail.
The calculation of the decay and energy loss
rates is performed in Sec.~\ref{sec3},
while results for the rates are presented and
bounds on the Lorentz-violating
parameters are derived in Sec.~\ref{sec4}.
Conclusions are reserved for Sec.~\ref{sec5}.

%
%
\section{Framework}
\label{sec2}

%
%
\subsection{Theoretical Basis and Assumptions}
\label{sec21}

The theoretical basis for the 
description of the Lorentz-violating states
is given by generalized Dirac equations,
which contain Lorentz-violating terms.
In typical cases [see, e.g., Eq.~(1) of Ref.~\cite{BeLe2012}],
these give rise to a dispersion relation
$E = | \vec p |\, v$ with a 
Lorentz-violating parameter $v > 1$. In the following, 
we parameterize the departure of $v$ from unity by 
setting $v = \sqrt{1+\delta}$.

If the virtuality of the incoming superluminal 
neutrino is large enough,
\begin{equation}
E^2 - \vec p^{\,2} \geq (2 m_e)^2 \,,
\end{equation}
then electron-positron pair production becomes possible.
This translates into an energy threshold
\begin{equation}
\label{EthLPCR}
E_{\rm th} = 
\frac{2 m_e}{\sqrt{\delta}} \,,
\qquad
\delta = v^2 - 1\,,
\end{equation}
for the onset of light lepton pair production,
or, lepton-pair Cerenkov radiation~\cite{CoGl1999,CoGl2011},
as manifest in the reaction
\begin{equation}
\label{LPCR}
\nu \to \nu + e^- + e^+ \,.
\end{equation}
Notice 
that Eq.~\eqref{EthLPCR} is correct in the small $\delta$ limit up 
to terms of relative order $\delta$. The exact expression 
for the energy threshold is $E_{\rm th} = 2 m_e \sqrt{1+\delta}/\sqrt{\delta}$.

\begin{figure} [t]
\begin{center}
\begin{minipage}{0.99\linewidth}
\begin{center}
\includegraphics[width=0.91\linewidth]{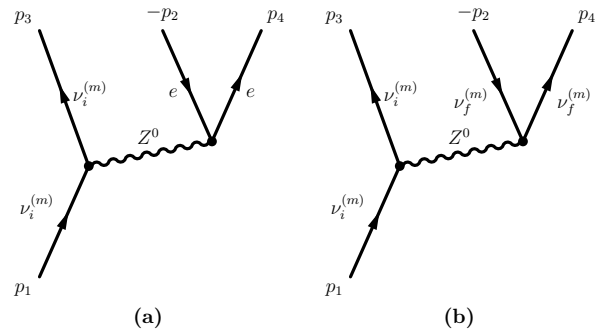}
\caption{\label{fig1}
The Feynman diagrams for the LPCR [Fig.~(a)]
and NPCR [Fig.~(b)] processes illustrate the 
exchange of a virtual $Z$ boson. They become kinematically 
possible for an incoming neutrino mass eigenstate
$| \nu^{(m)}_i \rangle$, which decays 
into a state of the same mass, but lower energy,
also labeled $| \nu^{(m)}_i \rangle$, 
and an electron-positron pair (LPCR)
or a neutrino-antineutrino pair (NPCR).}
\end{center}
\end{minipage}
\end{center}
\end{figure}

The process~\eqref{LPCR} gives rise to both a 
decay rate as well as an energy loss rate,
for any incoming superluminal neutrino,
as illustrated in Eqs.~(2) and~(3) of Ref.~\cite{CoGl2011}.
In choosing our convention for the $\delta$ parameter in Eq.~\eqref{EthLPCR},
according to $v \approx 1 + \delta/2$,
we are consistent with Refs.~\cite{CoGl1999,CoGl2011,BeLe2012,JeNaEh2017},
while Refs.~\cite{St2014,StSc2014,StEtAl2015} choose 
their $\delta$ parameters to be equal to just the 
difference $v-1$.
We here adhere to the conventions chosen in 
previous calculations of neutrino decay processes,
given in Refs.~\cite{CoGl2011,BeLe2012,JeNaEh2017}.
In Ref.~\cite{CoGl2011}, it was stated
that yet another process,
\begin{equation}
\label{NPCR}
\nu_i \to \nu_i + \nu_f + \overline \nu_f \,,
\end{equation}
might become kinematically possible
for superluminal neutrinos.
For other processes 
limiting superluminal models in the low-energy 
domain, we refer to Ref.~\cite{BiEtAl2011}.
Both processes in Eqs.~\eqref{LPCR} and~\eqref{NPCR}
involve the exchange of a $Z^0$ boson,
as is evident from the diagrammatic representation
in Fig.~\ref{fig1}.
The interesting feature of the process given in Eq.~\eqref{NPCR} 
is that the corresponding energy threshold, 
\begin{equation}
\label{EthNPCR}
E_{\rm th} = \frac{2 m_\nu}{\sqrt{\delta_i}} \,,
\qquad
\delta_i = v_i^2 - 1 \,,
\end{equation}
is smaller than~\eqref{EthLPCR} by 
at least six orders of magnitude
and can thus safely be ignored~\cite{StEtAl2015}.
Here, the parameter $\delta_i$ 
describes the Lorentz violation of the 
initial neutrino mass eigenstate as depicted in Fig.~\ref{fig1}.
We assume all $\delta$ parameters in this paper to be positive.
The parameter $m_\nu$ in Eq.~\eqref{EthNPCR}
denotes the mass of the ``final'' mass eigenstate 
denoted as $\nu_f = \nu_f^{(m)}$ in Fig.~\ref{fig1},
assuming a dispersion relation of the form
given below in Eq.~\eqref{disp_rel}.
(We shall comment on the suitability of the 
mass eigenstate basis for the calculation in the following.)
For reference, and somewhat 
pessimistically, we assume that 
$m_\nu$ is the largest among the 
masses of the neutrino mass eigenstates;
even under this assumption, the mass
$m_\nu$ entering the 
threshold~\eqref{EthNPCR}
is still safely smaller than $1 \, {\rm eV}$.
In the following, we shall 
concentrate on the high-energy region, where mass
terms in the dispersion relation~\eqref{disp_rel}
become irrelevant, and
pair production becomes possible for 
$\delta_i > \delta_f$, as an investigation of the 
available phase space for final states shows.
We thus investigate a possible 
additional relevance of neutrino-pair Cerenkov 
radiation, the process given in Eq.~\eqref{NPCR},
as an additional means of deriving
limits on Lorentz-violating parameters 
for superluminal neutrinos.

To this end, we assume that it is appropriate to 
generalize the dispersion relation 
$E = | \vec p| \, v$ to the following form
(see Ref.~\cite{Al2007} and Appendix~\ref{appa}),
\begin{equation}
\label{disp_rel}
E_k = \sqrt{ \vec p^{\,2} \, v_k^2 + m_k^2 \, v_k^4} \,,
\end{equation}
where we consider an (incoming or final) mass 
eigenstate $k=1,2,3$.
Here, the subscript $k=i$ denotes, simultaneously,
the initial state of the process, 
and equivalently a mass eigenstate,
while $k=f$ denotes the final state.
We know that the masses $m_i$ are nonvanishing and
different from each other,
so that it makes perfect sense
[see Eq.~\eqref{disp_rel}] to also 
consider slightly different 
Lorentz-violating parameters $v_i$,
even though their relative differences 
are very tightly constrained~\cite{CoGl1999,AaEtAl2010lorentz,%
AAEtAl2013boone}.

A difficulty arises.
Namely, because the weak-interaction Lagrangian
is flavor-diagonal,
neutrinos are always generated in 
flavor eigenstates.
However, if we assume the Lorentz-violating 
parameters to be different, 
then we need to consider the 
decay process in the basis of mass eigenstates.
Let us denote a mass eigenstate 
by the superscript $(m)$,
with eigenstates labeled as $k = 1,2,3$,
and a flavor eigenstate by the superscript $(f)$,
with an eigenstate labeled by the 
subscript $\ell = 1,2,3$.
The two are related by the PMNS matrix $U$, 
\begin{equation}
\nu^{(f)}_k = \sum U_{k \ell} \, \nu^{(m)}_\ell \,.
\end{equation}
For the coupling to a $Z^0$ boson,
the interaction Lagrangian is
\begin{equation}
\label{lag_escalafon1}
\calL = -\frac{g_w}{4 \, \cos\theta_W} \,
\sum_k \overline \nu^{(f)}_k \, \gamma^\mu (1 - \gamma^5) \,
\nu^{(f)}_k \, Z_\mu \,,
\end{equation}
where $Z_\mu$ denotes the $Z$ boson field,
and the $\gamma^\mu$ are the usual Dirac matrices,
while $\gamma^5$ is the fifth current Dirac matrix.
The weak coupling constant is $g_w$, and $\theta_W$ is the 
Weinberg angle. 
We can reformulate Eq.~\eqref{lag_escalafon1} as
%
%
\begin{equation}
\calL = -\frac{g_w}{4 \, \cos\theta_W} \,
\sum_{k,\ell,\ell'} U^+_{\ell k} \, U_{k \ell'} \,
\overline\nu^{(m)}_\ell \, \gamma^\mu (1 - \gamma^5) \,
\nu^{(m)}_{\ell'}  \, Z_\mu \,,
\end{equation}
where all summation indices cover the range $k,\ell,\ell' = 1,2,3$.
Assuming unitarity of the PMNS matrix,
$\sum_{k} U^+_{\ell k} \, U_{k \ell'} = \delta_{\ell \ell'}$,
one arrives at
\begin{equation}
\label{lag_escalafon2}
\calL = -\frac{g_w}{4 \, \cos\theta_W} \,
\sum_{\ell} 
\overline\nu^{(m)}_\ell \, \gamma^\mu (1 - \gamma^5) \,
\nu^{(m)}_{\ell}  \, Z_\mu \,,
\end{equation}
exhibiting diagonality in the mass eigenstate basis.
We can thus variously choose to evaluate the 
decay, and energy loss, rates, in the basis of 
flavor eigenstates, given in Eqs.~\eqref{lag_escalafon1},
or in the basis of mass eigenstates,
according to Eq.~\eqref{lag_escalafon2}.

This leaves open the choice of the
free Lagrangian for the neutrino sector.
In the following, we shall ignore the
neutrino mass term in Eq.~\eqref{disp_rel}
and write a Lagrangian which is applicable in the 
high-energy limit.
Following Ref.~\cite{BeLe2012}, we write it as
\begin{equation}
\label{lag_free_nu}
\calL = \sum_\ell \ii \, \overline \nu^{(m)}_\ell \,
\gamma^\mu \, (1-\gamma^5) \,
{\tilde g}_{\mu\nu}(v_\ell) \,
\partial^\nu \nu^{(m)}_\ell \,,
\end{equation}
where 
\begin{equation}
\label{tildeg}
{\tilde g}_{\mu\nu}(v_\ell) = {\rm diag}(1, -v_\ell, -v_\ell, -v_\ell) 
\end{equation}
is a Lorentz-violating ``metric'' describing the 
free propagation of the $\ell$th mass eigenstate.
We show in the following that the 
dispersion relation for the $\ell$th mass eigenstate,
implied by Eq.~\eqref{lag_free_nu},
is $E_\ell = |\vec p| \, v_\ell$.

Our aim here is to present a unified treatment of both
LPCR as well as NPCR in the Lorentz-violating sector.
The generalized interaction Lagrangian can be derived from the 
mass-basis interaction~\eqref{lag_escalafon2} 
upon considering the exchange of a $Z$ boson under kinematic 
conditions where 
the spatial momentum of the virtual exchange boson
can be neglected. In Sec.~2.3 of~\cite{JeNaEh2017}, it has 
recently been shown that this is the case,
for both tachyonic 
(Lorentz-conserving) as well as Lorentz-violating
superluminal neutrinos,
for surprisingly large incoming neutrino energies.
Its range of applicability covers the entire kinematic regime 
from zero neutrino energy up to the ``Big Bird'' energy
(Refs.~\cite{AaEtAl2013, AaEtAl2014}) of $2 \, {\rm PeV}$.
We thus write the interaction Lagrangian as
\begin{align}
\label{Lint}
\calL_{\rm int} =& \;
f_e \, \frac{G_F}{2 \sqrt{2}} \,
\overline\nu^{(m)}_i \, \gamma^\lambda \, (1 - \gamma^5) \, \nu^{(m)}_i 
\nonumber\\[0.1133ex]
& \; \times 
{\tilde g}_{\lambda\sigma}(v_{\rm int}) \;
\bar{\psi}_f \, \gamma^\sigma \, (c_V - c_A \, \gamma^5) \, \psi_f \,.
\end{align}
When written in this form, the interaction Lagrangian subsumes 
both the form assumed in Ref.~\cite{CoGl2011} (with $v_{\rm int} = 0$), 
as well as models I and II of Ref.~\cite{BeLe2012} (with $v_{\rm int} = 0$ 
and $v_{\rm int} = v_i$ respectively. 
On the occasion,
we note that the difference between the results 
of Ref.~\cite{CoGl2011} and model I of Ref.~\cite{BeLe2012} arises due to the 
different treatment of the spin sum for the superluminal neutrino, see 
Eqs.~\eqref{CGspinsum} and \eqref{BLspinsum} below. However, we can 
also keep $v_{\rm int}$ as an additional free parameter of the model.
As pointed out in Appendix~\ref{appb}, 
gauge invariance with respect 
to a subgroup of the $SU(2)_L \times U(1)_Y$ 
electroweak gauge symmetry group is maintained 
for $v_{\rm int} = v_i \, v_f$. This observation also explains the 
remark following Eq.~(4) of Ref.~\cite{BeLe2012},
where the authors refer to a ``gauge invariant'' (sic! with 
quotation marks) model.
In Ref.~\cite{BeLe2012}, the authors assume that $v_f = 1$.
In Eq.~\eqref{Lint}, $\psi_f = e$ or $\nu_f^{(m)}$ is the fermion 
field for either the electron or a neutrino mass eigenstate, while
\begin{equation}
f_e = \left\{ \begin{array}{ll}
1, & \qquad \psi_f = \nu^{(m)}_f \\
2, & \qquad \psi_f = e \\ 
\end{array} \right. \,,
\label{eq:fe}
\end{equation}
and (approximately)
\begin{equation}
(c_V, c_A) = \left\{ \begin{array}{ll}
(1,1) & \qquad \psi_f = \nu^{(m)}_f \\
(0,-\tfrac12), & \qquad \psi_f = e \\
\end{array} \right. \,.
\label{eq:cAcV}
\end{equation}

Summarizing, we describe the decay processes depicted in Fig.~\ref{fig1}
on the basis of a model which contains the following parameters:
\begin{itemize}
\item The parameter $\delta_i = v_i^2 -1$ is the Lorentz-violating parameter
for the initial, decaying particle state (comprising momenta
$p_1$ and $p_3$ in Fig.~\ref{fig1}).
\item The other Lorentz-violating 
parameter $\delta_f = v_f^2 -1$ describes 
the produced pair (comprising momenta
$p_2$ and $p_4$ in Fig.~\ref{fig1}).
\item We have the Lorentz-violating ``metric'' 
${\tilde g}_{\mu\nu}(v_\ell) = {\rm diag}(1, -v_\ell, -v_\ell, -v_\ell)$
that enters the Dirac equation describing the neutrino mass eigenstates.
(Here, $\ell$ can be the initial mass eigenstate $i$
or the final eigenstate $f$.)
For an electron-positron final state, 
we may set $v_\ell = 1$ at the end of the computation,
according to the kinematic assumptions made.
\item A further model-dependent Lorentz-violating ``metric''
${\tilde g}_{\mu\nu}(v_{\rm int})$ enters the 
interaction Lagrangian given in Eq.~\eqref{Lint} and has 
an additional free parameter, $v_{\rm int}$, where obviously 
$v_{\rm int} \simeq 1$. In order to keep the interpretation of 
all parameters in our model similar, we set 
$v_{\rm int} = \sqrt{1+\delta_{\rm int}}$.
\color{black}
As pointed out in Appendix~\ref{appb}, gauge invariance 
with respect to a subgroup of
$SU(2)_L \times U(1)_Y$ is maintained if we set
$\delta_{\rm int} = \delta_i + \delta_f$.
\color{black}
\end{itemize}
In order to keep the scope of the current investigation finite,
we shall concentrate on the following cases:
\begin{itemize}
\item The Cohen--Glashow model for the LPCR process~\cite{CoGl2011}
has $\delta_f = 0$, and replaces
${\tilde g}_{\mu\nu}(v_{\rm int}) \to g_{\mu\nu}$
in the interaction Lagrangian,
corresponding to $\delta_{\rm int} = 0$. Furthermore, the spin sum for 
a superluminal neutrino is assumed to take the following standard form:
\begin{equation}
\label{CGspinsum}
\sum_s \nu_{\ell,s} \otimes \bar{\nu}_{\ell,s} = p^\mu g_{\mu\nu} \gamma^\nu \,.
\end{equation}
For the NPCR process, one 
uses these prescriptions but keeps $\delta_f$ as a free parameter.
\item The model~I considered by Bezrukov and Lee~\cite{BeLe2012}
for the LPCR process 
also has $\delta_f = 0$, and ${\tilde g}_{\mu\nu}(v_{\rm int}) = g_{\lambda\sigma}$,
corresponding to $\delta_{\rm int} = 0$, however the spin sum for 
a superluminal neutrino takes the form
\begin{equation}
\label{BLspinsum}
\sum_s \nu_{\ell,s} \otimes \bar{\nu}_{\ell,s} = 
p^\mu {\tilde g}_{\mu\nu}(v_\ell) \gamma^\nu \,,
\end{equation}
as implied by the Dirac equation in Eq.~\eqref{eq:Dirac-SL}. For the NPCR 
process, one again has to keep $\delta_f$ as a free parameter.
\item The model~II considered by Bezrukov and Lee~\cite{BeLe2012}
for the LPCR process also has $\delta_f = 0$, and ${\tilde g}_{\mu\nu}(v_{\rm int}) = 
{\tilde g}_{\mu\nu}(v_i)$, which implies that $\delta_{\rm int} = \delta_i$. Spin sums 
for superluminal neutrinos are evaluated as in Eq.~\eqref{BLspinsum}. 
For the NPCR process, one again keeps $\delta_f$ as a free parameter.
Models~I and~II of Bezrukov and Lee~\cite{BeLe2012}
can thus be unified on the basis of the parameter $\delta_{\rm int}$,
which assumes the value $\delta_{\rm int} = 0$ for model~I and 
$\delta_{\rm int} = \delta_i$ for model~II.
If one would like to maintain
gauge invariance, within a subgroup of 
$SU(2) \times U(1)_Y$, and allow for a nonvanishing $\delta_f$,
then one should replace $\delta_{\rm int} = \delta_i + \delta_f$.
\item Expressed differently,
one can interpolate between models~I and~II of 
Bezrukov and Lee~\cite{BeLe2012} by considering 
$\delta_{\rm int}$  to be an additional free parameter of the model. 
\end{itemize}
All models considered use the ``metric''
given in Eq.~\eqref{tildeg} for the description of initial 
and final states of the incoming and outgoing neutrinos.

%
%
\subsection{Formalism and Models}
\label{sec22}

In order to study the LPCR and NPCR processes, 
$\nu_i \to \nu_i \psi_f\bar{\psi}_f$ ($\psi = e\,,\nu$) 
in a unified way, we follow the framework laid out 
in~\Refr{BeLe2012}, where the Lagrangian for a free 
superluminal neutrino field of flavor $i$ reads
\begin{equation}
\calL = \bar\psi_i \, 
\ii \, \gamma^\mu \, {\tilde g}_{\mu\nu}(v_i) \, 
\partial^\nu \, (1 - \gamma^5) \, \psi_i \,,
\label{eq:Lagrangian-SL}
\end{equation}
where the Lorentz violating ``metric" ${\tilde g}_{\mu\nu}(v_i)$ 
has been defined in Eq.~\eqref{tildeg}.
As outlined above, we work with neutrino mass eigenstates 
and suppress the superscript $(m)$ in the following.
Note that in comparison to 
Appendix~\ref{appa}, we explicitly exhibit the left-handed chirality 
projection in Eq.~\eqref{eq:Lagrangian-SL}, 
and suppress the mass term.
Crucially for what follows, the speed of the neutrino, $v_i = \sqrt{1+\del_i}$, 
is allowed to depend on the flavor. The Lagrangian in Eq.~\eqref{eq:Lagrangian-SL} 
leads to the following superluminal Dirac equation
\begin{equation}
\gamma^\mu {\tilde g}_{\mu\nu}(v_i) p^\nu \psi_i = 0
\label{eq:Dirac-SL}
\end{equation}
by the usual variational equation. Upon multiplication
from the right by the operator
$\gamma^\rho {\tilde g}_{\rho\sigma}(v_i) p^\nu \psi_i = 0$,
the superluminal Dirac equation~\eqref{eq:Dirac-SL} 
implies that
\begin{equation}
g^{\mu\nu} {\tilde g}_{\mu\alpha}(v_i) 
{\tilde g}_{\nu\beta}(v_i) p^\alpha p^\beta \psi_i = 0\,,
\end{equation}
which in turn leads to the desired superluminal dispersion relation
\begin{equation}
E^2 - | \vec k |\, v_i^2 = 0\,,
\label{eq:disprel-SL} 
\end{equation}
where we identify the components of the 
four-vector $p^\nu$ as $(E, \vec k)$.

Before moving on to computing the decay rate, we pause briefly to reinterpret 
the above formulae in a form which is convenient for later calculations. The key 
observation is that by introducing the time-like vector
\begin{equation}
t_\mu = (1,0,0,0) \,,
\end{equation}
(assumed to take this form in the ``laboratory" frame) we can rewrite the Lorentz 
violating ``metric" 
${\tilde g}_{\mu\nu}(v_i)$ as follows:
\begin{align}
{\tilde g}_{\mu\nu}(v_i) =& \;
v_i g _{\mu\nu} + (1-v_i) t_\mu t_\nu 
\nonumber\\[0.1133ex]
=& \; 
\color{black}
{\rm diag}(1, -v_i, -v_i, -v_i) \,.
\end{align}
{\color{black}
A remark is in order. In Ref.~\cite{CoKo2001},
a potentially necessary field redefinition
has been discussed, in order to ensure that
the fields are canonically normalized in a
Lorentz-violating scenario.
However, we note that, for the metric~\eqref{tildeg},
the field redefinition transformation $A$ outlined in
Eqs.~(4), (10) and (11) of Ref.~\cite{CoKo2001}
amounts to the unity transformation;
hence no further corrections are incurred.
This can be be seen as follows.
First, we notice that the metric 
${\tilde g}_{\mu\nu}(v_i)$ 
has the properties ${\tilde g}_{00}(v_i) = 1$ 
and ${\tilde g}_{0\nu}(v_i) = \delta^{0\nu}$,
and hence, no extra time derivative pieces
are introduced in the modified action or
modified Dirac equation.

In the notation of Ref.~\cite{CoKo2001},
our model corresponds to the parameter choice
\begin{equation}
c^{\mu\nu}_i = (v_i-1) \;
{\rm diag}(0,-1,-1,-1) \,.
\end{equation}
The spinor redefinition of Ref.~\cite{CoKo2001},
which otherwise eliminates the extra time
derivatives,
amounts to the transformation 
\begin{equation}
\psi = A \, \chi \,,
\qquad
A = 1 - \frac12 \, c_{\mu 0} \, \gamma^0 \, \gamma^\mu \,,
\end{equation}
but in our model $c_{\mu 0} = 0$, so that we simply have $A = 1$,
i.e., no spinor redefinition is needed.}

The superluminal Dirac equation in Eq.~\eqref{eq:Dirac-SL} 
takes the form
\begin{equation}
\gamma^\mu [v_i g_{\mu\nu} + (1-v_i) t_\mu t_\nu] p^\nu \psi_i = 
[v_i \slashed{p} + (1-v_i) (p\cdot t) \slashed{t}]\psi_i = 0 \,,
\end{equation}
where the slashed notation {\em always} has its usual meaning of 
$\slashed{a} = \gamma^\mu g_{\mu\nu} a^\nu$ for any four-vector $a$, 
with $g_{\mu\nu}$ the usual metric. 
Furthermore, the spin sum in Eq.~\eqref{BLspinsum} can be written as
\begin{equation}
\label{eq:spinsum-SL-t}
\sum_s \, \nu_{i,s} \otimes \bar \nu_{i,s} = 
v_i \slashed{p} + (1-v_i) (p\cdot t)\slashed{t} \,.
\end{equation}
Clearly, this differs from the standard spin sum of Eq.~\eqref{CGspinsum}, 
$\sum_s \, \nu_{i,s} \otimes \bar \nu_{i,s} = \slashed{p}$, by terms that are 
first order in $\delta_i$, as already 
pointed out in Ref.~\cite{BeLe2012}.

With the effective interaction Lagrangian given in Eq.~\eqref{Lint}, 
we are now ready to consider the emission of a fermion pair from a 
superluminal neutrino in the most general setup with fermion 
flavor-dependent speeds as well as a generic $v_{\rm int}$. 

%
%
\section{Calculation of emission rates}
\label{sec3}

%
%
\subsection{Matrix Elements}
\label{sec31}

Let us consider the process
\begin{equation}
\label{eq:n_npp}
\nu_i(p_1) \to \nu_i(p_3) + \bar{\psi}_f(p_2) + \psi_f(p_4)\,,
\end{equation}
where we have indicated in the parentheses the momentum 
assignment as in Fig.~\ref{fig1}. As discussed above, we work in 
the mass basis for the neutrinos, and we can specify the type of 
fermion $\psi$ at the very end of the calculation.

The transition matrix element for the process in Eq.~\eqref{eq:n_npp} 
can be computed from the effective interaction Lagrangian given in 
Eq.~\eqref{Lint}:
\begin{multline}
{\mathcal M} = f_e \frac{G_F}{2\sqrt{2}} 
\left[\bar{u}_i(p_3)\gamma^\lambda(1-\gamma^5)u_i(p_1)\right]
{\tilde g}_{\lambda\sigma}(v_{\rm int})
\\[0.1133ex]
\times 
\left[\bar{u}_f(p_4)(c_V \gamma^\sigma - c_A \gamma^\sigma \gamma^5)v_f(p_2)\right]\,.
\end{multline}
Summation over the spins of the final state particles and averaging over those of the 
initial ones can be performed using standard trace technology. However, 
one must remember that except for the Cohen-Glashow model, the spin sums for 
superluminal particles should be written as in \eqref{eq:spinsum-SL-t}. 
Allowing for $n_s$ 
spin states of the neutrino ($n_s=2$ in \Refr{CoGl2011} but $n_s = 1$ 
in \Refr{BeLe2012}), we find
\begin{equation}
\label{eq:SME}
\begin{split}
&
\frac{1}{n_s} \sum_{\mathrm{spins}} |{\mathcal M}|^2 =
\frac{1}{n_s} f_e^2 \frac{G_F^2}{8} 
\Tr[(v_i \slashed{p}_3 + (1-v_i)(p_3\cdot t) \slashed{t}) 
\\& \times
\gamma^\lambda(1-\gamma^5)
(v_i \slashed{p}_1 + (1-v_i)(p_1\cdot t) \slashed{t}) \gamma^\sigma(1-\gamma^5)]
\\ 
& \times
[v_{\rm int} g _{\lambda\rho} + (1-v_{\rm int}) t_\lambda t_\rho]
[v_{\rm int} g _{\sigma\tau} + (1-v_{\rm int}) t_\sigma t_\tau]
\\ 
& \times
\Tr[(v_f \slashed{p}_4 + (1-v_f)(p_4\cdot t) \slashed{t}) (c_V \gamma^\rho - c_A \gamma^\rho \gamma^5)
\\ 
& \times
(v_f \slashed{p}_2 + (1-v_f)(p_2\cdot t) \slashed{t}) 
(c_V \gamma^\tau - c_A \gamma^\tau \gamma^5)] \,.
\end{split}
\end{equation}
{\color{black}
Let us remark here, for absolute clarification, that the 
use of the convention $n_s = 1$ in \Refr{BeLe2012}
of course does not imply that neutrinos are treated
as scalar particles by Bezrukov and Lee \Refr{BeLe2012}, 
but only means that the authors assumed that all oncoming particles have
left-handed helicity.
If we were to use $n_s = 2$ in all calculations 
reported below for the models of Bezrukov and Lee \Refr{BeLe2012}, 
then the corresponding results would have to be divided by a 
factor two.}

The computation of traces and contractions is straightforward, although cumbersome. 
The final result is somewhat long and will not be exhibited here, 
but the general structure 
is the following. The squared matrix element 
can be written as a linear combination of ten 
different kinematic structures, all of mass dimension four:
\begin{equation}
\begin{split}
& \frac{1}{n_s} \sum_{\mathrm{spins}} |{\mathcal M}|^2 
= \frac{1}{n_s} f_e^2 \frac{G_F^2}{8} \bigg[
c_1 (p_1\cdot p_2) (p_3\cdot p_4)
\\&
+ c_2 (p_1\cdot p_3) (p_2\cdot p_4)
+ c_3 (p_1\cdot p_4) (p_2\cdot p_3)
\\&
+ c_4 (p_1\cdot p_2) (p_3\cdot t) (p_4\cdot t)
+ c_5 (p_1\cdot p_3) (p_2\cdot t) (p_4\cdot t)
\\&
+ c_6 (p_1\cdot p_4) (p_2\cdot t) (p_3\cdot t)
+ c_7 (p_2\cdot p_3) (p_1\cdot t) (p_4\cdot t)
\\&
+ c_8 (p_2\cdot p_4) (p_1\cdot t) (p_3\cdot t)
+ c_9 (p_3\cdot p_4) (p_1\cdot t) (p_2\cdot t)
\\&
+ c_{10} (p_1\cdot t) (p_2\cdot t) (p_3\cdot t) (p_4\cdot t)
\bigg]\,.
\end{split}
\label{SME}
\end{equation}
Here the $c_i$ ($i=1,\ldots, 10$) are numerical coefficients that depend 
on $v_i$, $v_f$, $v_{\rm int}$ as well as $c_A$, $c_V$ and whether 
Eq.~\eqref{CGspinsum} or Eq.~\eqref{BLspinsum} is used to evaluate 
the spin sum for superluminal particles.

%
%
\subsection{Phase--Space Integration}
\label{sec32}

Now we are ready to compute the fermion emission rate, which 
essentially amounts to integrating the squared matrix element in 
Eq.~\eqref{SME} over the available phase space:
\begin{equation}
\label{Gamma}
\Gamma = \frac{1}{2E_1} \int \dd \phi_3(p_2,p_3,p_4;p_1) 
\frac{1}{n_s} \sum_{\mathrm{spins}} |{\mathcal M}|^2\,.
\end{equation}
A remark is in order.
Namely, in Ref.~\cite{CoKo2001},
small deviations of the formulas for the 
calculation of cross sections
in Lorentz-violating theories, 
from those in Lorentz-invariant theories, 
have been derived.
A closer inspection of Ref.~\cite{CoKo2001}
reveals that the correction
terms to the flux factors are of relative order 
of the $\delta$ parameters 
of the oncoming particles (assuming that the deviations
from the speed of light are small,
$\delta \ll 1$).
{\color{black}
The additional corrections would thus modify the results given 
in Sec.~\ref{sec41} at higher order in the 
expansion parameters $\delta$,
which we consider the terms to be numerically small.
In addition, a closer inspection reveals that the
tiny modifications necessary for the 
cross sections in Lorentz-violating theories,
can be traced to
flux factor which involves the relative velocity of the in-
coming particles, as shown in Ref.~\cite{CoKo2001}. 
However, the
formula for the decay rate does not involve a flux factor,
and hence, Eq.~\eqref{Gamma} does not require further modifications.}
The same approach has been taken
in Refs.~\cite{CoGl2011,BeLe2012}.

We evaluate the integral by first using the splitting relation~\cite{ByKa1973} 
(in the ``laboratory'' frame)  to write
\begin{multline}
\Gamma = \frac{1}{2E_1} \int_{M^2_{\mathrm{min}}}^{M^2_{\mathrm{max}}} \frac{\dd M^2}{2\pi}
\dd \phi_2(p_3,p_{24};p_1) \, \dd \phi_2(p_2,p_4;p_{24}) 
\\
\times \frac{1}{n_s} \sum_{\mathrm{spins}} |{\mathcal M}|^2 \,.
\end{multline}
The product of the two factors, $\dd \phi_2(p_3,p_{24};p_1)$ 
times $\dd \phi_2(p_2,p_4;p_{24})$, displays the kinematics of the process, 
in two steps, $p_1 \to p_3 + p_{24}$ and $p_{24} \to p_2 + p_4$.
Although the splitting relation is well-known~\cite{ByKa1973} 
and has been used by us in the analysis of 
the three-jet phase space in quantum chromodynamics
in next-to-next-to-leading order
[see Eq.~(2.9) of Ref.~\cite{So2013} and Eq.~(2.12) of Ref.~\cite{DDSoTr2013}],
we give a quick derivation 
here to emphasize the fact that it holds without change also for superluminal 
momenta. We start with the definition of the three-particle phase space 
for the final state, $\dd \phi_3$,
\begin{multline}
\dd \phi_3(p_2,p_3,p_4;p_1) =
\frac{\dd^4 p_2}{(2\pi)^3} \del_{+}(p_2^2 - \del_f k_2^2)
\\
\times
\frac{\dd^4 p_3}{(2\pi)^3} \del_{+}(p_3^2 - \del_i k_3^2)
\frac{\dd^4 p_4}{(2\pi)^3} \del_{+}(p_4^2 - \del_f k_4^2)
\\
\times
(2\pi)^4 \del^{(4)}(p_1-p_2-p_3-p_4)\,,
\end{multline}
\color{black}
where we recall the definition of the 
components of the four-vector $p^\nu$ as $p^\nu = (E, \vec k)$,
from Eq.~\eqref{eq:disprel-SL}, and note that 
\begin{multline}
g^{\mu\nu} \tilde g_{\mu\alpha} \, \tilde g_{\mu\beta} \,
p^\alpha \, p^\beta = E^2 - v^2 \vec k^2 \\
= g_{\mu\nu} p^\mu \, p^\nu - (v^2 - 1) \, \vec k^2
= p^2 - \del \; k^2 \,.
\end{multline}
Furthermore,
\begin{equation}
\del_{+}(p^2 - \del \; k^2) =
\del_{+}(E^2 - v^2 \, k^2) =
\frac{1}{2 E} \, \delta(E - v \, | k |)
\end{equation}
identifies a Dirac-$\delta$ which is nonzero 
only for positive values of $E = p^0$.
\color{black}
We now insert a factor of one, written in the following form,
\begin{equation}
\label{introM}
1 = \int \dd^4 p_{24}\, \del^{(4)}(p_{24} - p_2 - p_4)
\int \dd M^2 \del_{+}(p_{24}^2 - M^2)\,.
\end{equation}
The second delta function implies that $p_{24}$ is a generic massive momentum. 
This is appropriate, since in general the sum of two superluminal momenta will not satisfy 
the superluminal dispersion relation in the ``laboratory'' frame. Thus we find
\begin{equation}
\begin{split}
& \dd \phi_3(p_2,p_3,p_4;p_1) =
\int \frac{\dd M^2}{2\pi}
\frac{\dd^4 {p_3}}{(2\pi)^3} \del_{+}(p_3^2 - \del_i k_3^2)
\\ 
& \times \frac{\dd^4 p_{24}}{(2\pi)^3} \del_{+}(p_{24}^2 - M^2)
(2\pi)^4 \del^{(4)}(p_1-p_3-p_{24})
\\ 
& \times
\frac{\dd^4 p_2}{(2\pi)^3} \del_{+}(p_2^2 - \del_f k_2^2)
\frac{\dd^4 p_4}{(2\pi)^3} \del_{+}(p_4^2 - \del_f k_4^2)
\\ 
& \times
(2\pi)^4 \del^{(4)}(p_{24}-p_2-p_4)
\\ 
& 
= \int \frac{\dd M^2}{2\pi} \dd \phi_2(p_3,p_{24};p_1) \dd \phi_2(p_2,p_4;p_{24})\,,
\end{split}
\end{equation}
i.e., the result is formally identical to the usual case with no superluminal particles.

We must now work out the limits of integration of the $M^2$ integral. To do this, let us 
recall that by momentum conservation ($p_1^\mu = p_2^\mu + p_3^\mu + p_4^\mu$) 
we have
\begin{subequations}
\begin{align}
p_{24}^\mu =& \; p_2^\mu + p_4^\mu = p_1^\mu - p_3^\mu \,,
\\[0.1133ex]
M^2 =& \; p_{24}^2 = (p_1-p_3)^2\,.
\end{align}
\end{subequations}
However, using the superluminal dispersion relation, we find
\begin{multline}
(p_1-p_3)^2 = p_1^2+p_3^2 - 2p_1\cdot p_3 = \del_i \avp{1}^2 + \del_i \avp{3}^2 \\
 - 2(1+\del_i) \avp{1} \avp{3} 
+ 2 \avp{1} \avp{3} \cos\theta_{13}\,,
\end{multline}
where $\theta_{13}$ is the angle between the three-momenta $\vp_1$ and $\vp_3$. 
Clearly, the expression above attains its maximum when $\cos\theta_{13} = 1$, so
\begin{equation} 
M^2_{\mathrm{max}} = \del_i (\avp{1} - \avp{3})^2\,.
\end{equation}
On the other hand, we have also
\begin{multline}
M^2 = (p_2 + p_4)^2 = p_2^2+p_4^2 + 2p_2\cdot p_4 = 
\del_f \avp{2}^2 + \del_f \avp{4}^2 \\
+ 2(1+\del_f) \avp{2} \avp{4} - 2 \avp{2} \avp{4} \cos\theta_{24}\,,
\end{multline}
where $\theta_{24}$ is now the angle between the three-momenta $\vp_2$ and $\vp_4$. 
Obviously, the expression above is minimal when $\cos\theta_{24} = 1$, so
\begin{equation} 
M^2_{\mathrm{min}} = \del_f (\avp{2} + \avp{4})^2\,.
\end{equation}
However, energy conservation in the laboratory frame simply reads
\begin{equation} 
\sqrt{1+\del_i} \avp{1} = \sqrt{1+\del_f} \avp{2} + \sqrt{1+\del_i} \avp{3} + \sqrt{1+\del_f} \avp{4}\,,
\end{equation}
implying 
\begin{equation}
\avp{2} + \avp{4} = \frac{\sqrt{1+\del_i}}{\sqrt{1+\del_f}} (\avp{1} - \avp{3})\,,
\end{equation}
hence finally
\begin{subequations}
\label{M2maxmin}
\begin{align}
M^2_{\mathrm{max}} =& \;
\del_i (\avp{1} - \avp{3})^2 \,,
\label{eq:Mmax}
\\[0.1133ex]
M^2_{\mathrm{min}} =& \;
\del_f \frac{1+\del_i}{1+\del_f} (\avp{1} -\avp{3})^2\,.
\label{eq:Mmin}
\end{align}
\end{subequations}
In particular, for $\delta_i = \delta_f$ the limits coincide and hence there is no 
phase space for the decay. (For $\delta_i < \delta_f$ the maximum allowed 
value of $M^2$ would be lower than the minimum allowed value.)

The utility of the splitting relation lies in the fact that we can 
first perform the integration of the squared matrix element over 
the momenta of the two outgoing ``slow'' particles
(where ``slow'' refers to a potentially still superluminal particle
with $\delta_f < \delta_i$). These are $p_2$ and $p_4$. To do this, 
we must evaluate the tensor integral (for $p_2$ and $p_4$ both superluminal)
\begin{equation}
J^{\mu\nu}(p_{24}) = \int \dd \phi_2(p_2,p_4;p_{24}) p_{2}^{\mu} p_{4}^{\nu}\,.
\label{eq:Jmunu}
\end{equation}
This can be computed using the following explicit parameterization 
of the two-particle phase space $\dd \phi_2(p_2,p_4;p_{24})$, valid 
when both $p_2$ and $p_4$ are superluminal with speed $v_f$:
\begin{multline}
\int \dd \phi_2(p_2,p_4;p_{24}) = \frac{1}{8\pi} 
\int\limits_{\avp{2}_{\mathrm{min}}}^{\avp{2}_{\mathrm{max}}} 
\frac{\dd \avp{2} \dd(\cos\tht)}{(1+\del_f)^{3/2} \avp{24}} 
\\
\times \delta\left[\cos\tht - \frac{2 \sqrt{1+\del_f}E_{24} \avp{2} + \del_f \avp{24}^2 - M^2}
{2 (1+\del_f)\avp{24}\avp{2}}\right] \,,
\label{eq:PS2_24}
\end{multline}
where the limits of integration read
\begin{subequations}
\begin{align}
\avp{2}_{\mathrm{max}} &= \frac{E_{24} + \sqrt{1+\del_f}\avp{24}}{2\sqrt{1+\del_f}}\,,
\\[0.1133ex]
\avp{2}_{\mathrm{min}} &= \frac{E_{24} - \sqrt{1+\del_f}\avp{24}}{2\sqrt{1+\del_f}}\,.
\end{align}
\end{subequations}
Then the tensor integral can be computed (e.g., component by component) 
and we find
\begin{multline}
J^{\mu\nu}(p_{24}) = \frac{1}{12}
\left[\frac{E_{24}^2 - (1+\del_f)\avp{24}^2}{1+\del_f} g^{\mu\nu} 
+ 2 p_{24}^\mu p_{24}^\nu
\right.
\\
\left.
+ \del_f \frac{E_{24}^2 - (1+\del_f)\avp{24}^2}{1+\del_f} t^\mu t^\nu\right]V_2^{2\mathrm{SL}}(\del_f)\,,
\label{eq:Jmunu-2SL}
\end{multline}
where $V_2^{2\mathrm{SL}}(\del_f)$ is the volume of the 
two-body phase space when both particles are superluminal and 
have a speed of $v_f$, 
\begin{equation}
V_2^{2\mathrm{SL}}(\del_f) = \frac{1}{8\pi(1+\del_f)^{3/2}}\,.
\end{equation}

\begin{figure} [t]
\begin{center}
\begin{minipage}{0.99\linewidth}
\begin{center}
\includegraphics[width=0.91\linewidth]{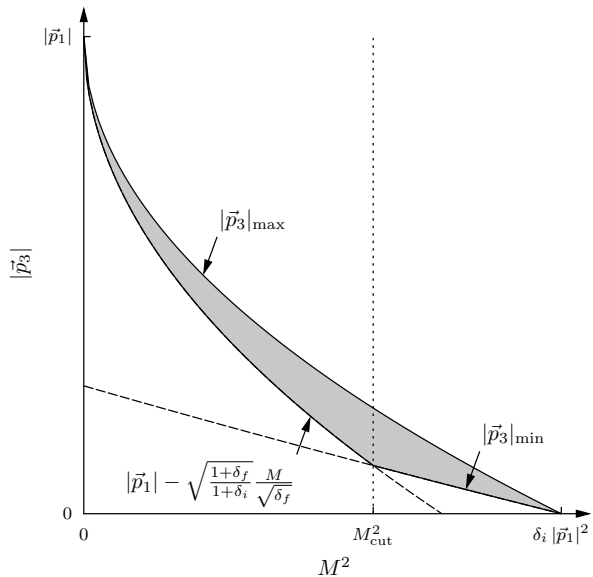}
\caption{\label{fig:PS}
The physical region (shaded) in the $(M^2,\avp{3})$ plane for ``slow'' 
superluminal neutrino pair emission from a ``fast'' superluminal neutrino. 
$\avp{3}_{\mathrm{max}}$ and $\avp{3}_{\mathrm{min}}$ are defined in 
Eq.~\eqref{eq:p3limits}, while $M^2_{\rm cut}$ is given in Eq.~\eqref{eq:M2cut}.
The variable $M^2$ is introduced in Eq.~\eqref{introM}
[see also Eq.~\eqref{M2maxmin}].}
\end{center}
\end{minipage}
\end{center}
\end{figure}

Substituting this, we are left with the integral over the two-body phase 
space $\dd \phi_2(p_3,p_{24};p_1)$ and over $\dd M^2$. The integration 
is essentially straightforward, however, care must be taken to properly 
identify the limits of integration in all variables. In particular, the phase 
space for one superluminal momentum with speed $v_i$ and one massive 
momentum with mass $M^2$ can be written in the following explicit form:
\begin{multline}
\int \dd \phi_2(p_3,p_{24};p_1) = \frac{1}{8\pi} 
\int_{\avp{3}_{\mathrm{min}}}^{\avp{3}_{\mathrm{max}}} \frac{\dd \avp{3} \dd(\cos\tht)}{\sqrt{1+\del_i} \avp{1}}
\\
\times 
\delta\left[\cos\tht - \frac{2(1+\del_i) \avp{1}\avp{3} + M^2 - \del_i \avp{1}^2 - \del_i \avp{3}^2}{2 \avp{1} \avp{3}}\right] \,,
\label{eq:PS2_3_24}
\end{multline}
where $\tht$ is the angle between the incoming and outgoing three-momentum of 
the ``fast'' superluminal neutrino. The limits of integration are
\begin{subequations}
\label{eq:p3limits}
\begin{align}
\avp{3}_{\mathrm{max}} =& \; \avp{1} - \frac{M}{\sqrt{\del_i}} \,,
\\[0.1133ex]
\avp{3}_{\mathrm{min}} =& \;
\frac{(2+\del_i) \avp{1} - \sqrt{4(1+\del_i) \avp{1}^2 + \del_i M^2}}{\del_i}\,.
\end{align}
\end{subequations}
However, phase space is also constrained by the limits of the $\dd M^2$ 
integration, given in Eqs.~\eqref{eq:Mmax} and \eqref{eq:Mmin}.
As remarked earlier, if $\del_i = \del_f$, then $M^2_{\mathrm{max}}$ 
and $M^2_{\mathrm{min}}$ coincide and there is no phase space for 
the decay. A careful but straightforward analysis establishes that the 
true region of integration corresponds to
\begin{equation}
\avp{3}_{\mathrm{min}} \le \avp{3} \le \avp{3}_{\mathrm{max}}
\end{equation}
only for 
\begin{equation}
M^2_{\mathrm{cut}} \le M^2  \le \del_i \avp{1}^2\,,
\end{equation}
while on the other hand
\begin{equation}
\avp{1} - \sqrt{\frac{1+\del_f}{1+\del_i}} \frac{M}{\sqrt{\del_f}} \le \avp{3} \le 
\avp{3}_{\mathrm{max}}\,.
\end{equation}
if
\begin{equation}
0 \le M^2 \le M^2_{\mathrm{cut}}\,,
\end{equation}
where 
\begin{equation}
M^2_{\mathrm{cut}} = 4\del_f(1+\del_i) \frac{\left(\sqrt{1+\del_i} - \sqrt{1+\del_f}\,\right)^2}{(\del_i - \del_f)^2} \avp{1}^2\,.
\label{eq:M2cut}
\end{equation}
Thus, the integrals can be performed most easily by splitting 
the $\dd M^2$ integration at $M^2_{\mathrm{cut}}$. The physical 
region in the $(M^2,\avp{3})$ plane is shown in Fig.~\ref{fig:PS}.

Before presenting our results, let us comment briefly on the computation of the
differential energy loss, given by~\cite{JeNaEh2017},
\begin{equation} \frac{\dd E_1}{\dd x} = -\int \dd E'_3 (E_1 - E'_3) \frac{\dd
\Gamma}{\dd E'_3}\,.  \label{eq:dEdx} \end{equation}
This formula first and foremost applies to the energy loss per time, which
translates into an energy loss per traveled distance for particle velocities
near the speed of light ($c=1$ in our conventions). The differential decay rate
is then simply 
\begin{equation} \frac{\dd \Gamma}{\dd E'_3} = \frac{1}{2E_1} \int
\phi_3(p_2,p_3,p_4;p_1) \frac12 \sum_{\mathrm{spins}} |{\mathcal M}|^2 \del(E_3
- E'_3)\,.  \end{equation}
When evaluating this expression, we need to keep in mind that in our
parameterization of phase space (see Eq.~\eqref{eq:PS2_3_24}), $E_3$ is assumed
to be expressed as a function of $\avp{3}$, i.e., $E_3 =
\sqrt{1+\del_i}\,\avp{3}$.  Hence we must replace $E_3$ in the above equation
by $\sqrt{1+\del_i}\,\avp{3}$. In particular, the Dirac delta function in the
above equation becomes $\del(\sqrt{1+\del_i}\,\avp{3} - E'_3)$.  We can then
perform the $E'_3$ integration in \eqref{eq:dEdx} with this delta function
first, so
\begin{align}
\frac{\dd E_1}{\dd x} =& \; -\int \dd E'_3 (E_1 - E'_3) \frac{\dd \Gamma}{\dd E'_3}
\nonumber\\[0.1133ex]
=& \;
-\int \dd E'_3 (E_1 - E'_3) \frac{1}{2E_1} \int \dd \phi_3(p_2,p_3,p_4;p_1)
\nonumber\\[0.1133ex]
& \; \times \frac12 \sum_{\mathrm{spins}} |{\mathcal M}|^2 \del(E_3 - E'_3)
\nonumber\\[0.1133ex]
=& \;
-\frac{1}{2E_1} 
\int  \dd \phi_3(p_2,p_3,p_4;p_1) 
\nonumber\\[0.1133ex]
& \; \times
\sqrt{1+\del_i}(\avp{1} - \avp{3})
\frac12 \sum_{\mathrm{spins}} |{\mathcal M}|^2\,,
\label{eq:dEdx-2}
\end{align}
where we have used $E_1 = \sqrt{1+\del_i}\,\avp{1}$ and
$E_3 = \sqrt{1+\del_i}\,\avp{3}$.
We can perform the integration over the phase space exactly 
as before, in fact, the calculation can be mapped onto the one 
for the decay rate by simply replacing 
$|{\mathcal M}|^2 \to -\sqrt{1+\del_i}\,(\avp{1} - \avp{3})|{\mathcal M}|^2$.

%
%
\section{Decay and Energy Loss Rates}
\label{sec4}

%
%
\subsection{Analytic Results}
\label{sec41}

The complete results for the decay rate and differential 
energy loss in full generality are quite cumbersome.
Hence, inspired by the approach of Ref.~\cite{CoGl2011}, we 
present here only the leading order results in 
the small quantities $\del$, 
but keep all terms of third order in the $\del$ parameters. 
More precisely, we assume 
that $\del_i \sim \del_f \sim \del_{\rm int}$ are all of the 
same order of magnitude and perform an expansion to 
the first non-vanishing order.

First, we present a general result, namely, for  
the total rate for the process $\nu_i \to \nu_i \psi_f \bar{\psi}_f$ 
(where $\psi_f$ and $\bar\psi_f$ can be any of the 
discussed outgoing fermions and anti-fermions, i.e., 
an electron-positron or a neutrino-antineutrino pair).
This general result can be written in the following form,
for a Lorentz-violating parameter $\delta_i$ 
of the incoming neutrino, and $\delta_f$ for the 
outgoing particle-antiparticle pair,
\begin{equation}
\label{eq:Gamma-general}
\begin{split}
&\Gamma_{\nu_i \to \nu_i \psi_f \bar{\psi}_f} = \frac{G_F^2 k_1^5}{192\pi^3}
	f_e^2 \frac{c_V^2+c_A^2}{420 n_s} (\delta_i-\delta_f) 
\\ &\quad\times
	\bigg[(60 - 43 \sigma_i)(\delta_i-\delta_f)^2
\\ & \quad\quad+
	2(50 - 32\sigma_i - 25\sigma_f + 7 \sigma_i \sigma_f)
	(\delta_i-\delta_f) \delta_f
\\ & \quad\quad+
	7(4 - 3\sigma_i - 3\sigma_f + 2 \sigma_i \sigma_f)
	\delta_f^2
	+ 
	7 \delta_{\rm int}^2\bigg]\,,
\end{split}
\end{equation}
\color{black}
where we notice the parameter $n_s$ for the number
of active spin states in the denominator,
\color{black}
Furthermore, $\sigma_i$ and $\sigma_f$ are zero or one depending on whether the 
Cohen-Glashow [see Ref.~\cite{CoGl2011} and Eq.~\eqref{CGspinsum}]
or the Bezrukov-Lee [see Ref.~\cite{BeLe2012} and Eq.~\eqref{eq:spinsum-SL-t}]
prescription is adopted for the sum over 
spins of the initial and final particles:
\begin{equation}
\sigma_i = \left\{ \begin{array}{ll}
0, & \quad \mbox{CG spin sum for $\nu_i$\,,}\\
1, & \quad \mbox{BL spin sum for $\nu_i$\,,}\\ 
\end{array} \right.
\end{equation}
and
\begin{equation}
\sigma_f = \left\{ \begin{array}{ll}
0, & \quad \mbox{CG spin sum for $\psi_f$\,,}\\
1, & \quad \mbox{BL spin sum for $\psi_f$\,.}\\ 
\end{array} \right.
\end{equation}
Furthermore, $\delta_{\rm int} = v_{\rm int}^2 - 1$ 
is the Lorentz-violating parameter for the 
metric used in the interaction Lagrangian~\eqref{Lint},
and $n_s$ is the number of available spin states
assumed in a particular model.
For the differential energy loss, we find
\begin{equation}
\label{eq:dEdx-general}
\begin{split}
&\frac{\dd E_{\nu_i \to \nu_i \psi_f \bar{\psi}_f}}{\dd x} 
= -\frac{G_F^2 k_1^6}{192\pi^3} 
f_e^2 \frac{c_V^2+c_A^2}{672 n_s} (\delta_i-\delta_f) 
\\ &\quad\times
\bigg[(75 - 53 \sigma_i)(\delta_i-\delta_f)^2
\\ & \quad\quad+
(122 - 77\sigma_i - 61\sigma_f + 16 \sigma_i \sigma_f)
(\delta_i-\delta_f) \delta_f
\\ & \quad\quad+
8(4 - 3\sigma_i - 3\sigma_f + 2 \sigma_i \sigma_f) \delta_f^2
+ 8 \delta_{\rm int}^2\bigg]\,,
\end{split}
\end{equation}
with $\sigma_i$ and $\sigma_f$ as above. Using the formulas in 
Eqs.~\eqref{eq:Gamma-general} and \eqref{eq:dEdx-general}, 
we can rederive the results of Refs.~\cite{CoGl2011,BeLe2012} 
by an appropriate choice of the various parameters,
as discussed in Sec.~\ref{sec2}.
It is interesting to observe that in Eq.~\eqref{eq:dEdx-general},
terms proportional to $c_V\,c_A$ vanish. This 
conclusion can be supported by a detailed analysis 
of the Dirac algebra of the transition 
currents and phase-space integrals: Namely, conceivable 
contributions proportional to $c_V\,c_A$ would 
be multiplied by an antisymmetric Dirac structure
(in the indices of the outgoing pair), multiplied
by a symmetric phase-space integral, and hence, they vanish.

We now proceed to the 
indication of the results for 
the superluminal models discussed here.
Because of a certain multitude 
of models discussed here, let us anticipate, 
for the convenience of the reader, the following 
conventions:
\begin{itemize}
\item 
The $a$ and $a'$ coefficients
given below in Eq.~\eqref{acoeff} refer to the vacuum pair emission,
or charged-lepton-pair Cerenkov radiation (LPCR),
with $a$ entering the formula for the 
decay rate, while $a'$ enters the formula for the 
energy loss rate.
\item 
The $b$ and $b'$ coefficients
given below in Eqs.~\eqref{resbCG} and~\eqref{resbBL}
refer to the neutrino-splitting 
or neutrino-pair Cerenkov radiation (NPCR),
with $b$ entering the formula for the 
decay rate, while $b'$ enters the formula for the 
energy loss rate.
\item Coefficients with a subscript CG refer
to the Cohen--Glashow model~\cite{CoGl2011}, 
which assumes the spin sum~\eqref{CGspinsum}
and has $n_s = 2$.
\item Coefficients with a subscript BL refer
to the Bezrukov--Lee  model~\cite{BeLe2012}, 
which assumes the spin sum~\eqref{BLspinsum}
and has $n_s = 1$.
\item The parameter $\delta_{\rm int}$ 
enters the effective interaction
Lagrangian~\eqref{Lint}.
According to the discussion in 
Appendices~\ref{appa} and~\ref{appb}
(see also Ref.~\cite{JeNaSo2019}),
\color{black} a restricted gauge structure
(with a reduced symmetry group)
of the electroweak interaction is preserved for 
$\delta_{\rm int} = \delta_i + \delta_e$
(LPCR process) and
$\delta_{\rm int} = \delta_i + \delta_f$
(NPCR process).
Here, the parameters $\delta_i$, $\delta_e$ and 
$\delta_f$ are measured with respect to the 
speed of light (see also Ref.~\cite{CoGl1997}).
\color{black}
\end{itemize}

Starting with the case $\psi_f = e$ (LPCR process), 
we recover the models of Cohen and Glashow (using the 
standard spin sum of Eq.~\eqref{CGspinsum}, i.e., $\sigma_i=\sigma_f=0$) 
of \Refr{CoGl2011} and of Bezrukov and Lee [using the superluminal spin 
sum of Eq.~\eqref{BLspinsum}, i.e., $\sigma_i=\sigma_f=1$]
of \Refr{BeLe2012}. We obtain for $\delta_f = \delta_e = 0$,
\begin{align}
\label{resLPCR}
\Gamma_{\nu_i \to \nu_i e^- e^+} &= a \frac{G_F^2}{192\pi^3} k_1^5\,,
\\
\frac{\dd E_{\nu_i \to \nu_i e^- e^+}}{\dd x} &= -a' \frac{G_F^2}{192\pi^3} k_1^6\,,
\end{align}
with
\begin{subequations}
\label{resa}
\begin{align}
a_{\mathrm{CG}} =& \; \frac{1}{14} \del_i^3 \,, &
\qquad
a'_{\mathrm{CG}} &= \frac{25}{448} \del_i^3\,,
\\[0.1133ex]
\label{coeffBLI}
a_{\mathrm{BL,I}} =& \; \frac{17}{420} \del_i^3 \,, &
\qquad
a'_{\mathrm{BL,I}} &=  \frac{11}{336} \del_i^3\,,
\\[0.1133ex]
\label{coeffBLII}
a_{\mathrm{BL,II}} =& \; \frac{2}{35} \del_i^3 \,, &
\qquad
a'_{\mathrm{BL,II}} & = \frac{5}{112} \del_i^3\,,
\end{align}
\end{subequations}
for the model of Cohen and Glashow~\cite{CoGl2011},
and models I and II of Bezrukov and Lee~\cite{BeLe2012}. 
{\color{black}
In regard to the models
of Bezrukov and Lee [see also Eq.~\eqref{eq:Gamma-general}], 
we recall that they correspond to
{\em (i)} using superluminal spin sums for all particles, so 
$\sigma_i = \sigma_f = 1$, {\em (ii)} setting $n_s = 1$
for the number of spin states
for the neutrino, {\em (iii)} considering the emitted electron to
be Lorentz-invariant, hence $\delta_f = 0$, 
and {\em (iv)} using $\delta_{\rm int} = 0$
for model I and $\delta_{\rm int} = \delta_i$ for model II. 
In this case too, $\sin^2(\theta_W) = 1/4$ is used.}
We thus confirm all 
known results from Refs.~\cite{CoGl2011,BeLe2012}.

Under the inclusion of a conceivably nonvanishing
parameter $\delta_e \neq 0$, 
the results generalize to the form,
\begin{align}
\label{acoeff}
a_{\mathrm{CG}} =& \; \frac{1}{14} (\del_i - \del_e) 
\left[(\del_i - \del_e)^2 + \frac{5}{3} \del_e (\del_i - \del_e) 
+ \frac{7}{15}\del_e^2\right]\,,
\\
a'_{\mathrm{CG}} =& \; \frac{25}{448} (\del_i - \del_e) 
\left[(\del_i - \del_e)^2 \right.
\nonumber\\[0.1133ex]
& \; \left. 
+ \frac{112}{75} \del_e (\del_i - \del_e) + \frac{32}{75}\del_e^2\right]\,,
\\[0.1133ex]
\label{aBL}
a_{\mathrm{BL}} =& \; \frac{17}{420} (\del_i - \del_e) 
\left[(\del_i - \del_e)^2 + \frac{7}{17}\del_{\rm int}^2\right]\,,
\\
\label{aprimeBL}
a'_{\mathrm{BL}} =& \; \frac{11}{336} (\del_i - \del_e)
\left[(\del_i - \del_e)^2 + \frac{4}{11}\del_{\rm int}^2\right]\,,
\end{align}
where one would set $\delta_{\rm int} = \delta_i + \delta_e$
in a gauge-invariant model (see the Appendices).

{\color{black} Just to achieve full clarification,
we remark that the interpolation between
$a_{\mathrm{BL,I}} \to a_{\mathrm{BL,II}}$
[see Eqs.~\eqref{coeffBLI} and~\eqref{coeffBLII}],
consistent with Eq.~\eqref{aBL}, 
can be traced to the algebraic identity
\begin{equation}
\frac{17}{420} \times \left( 1 + 
\frac{7}{17} \right) = \frac{2}{35}
\qquad
\mbox{[$a_{\mathrm{BL,I}} \to a_{\mathrm{BL,II}}$] \,.}
\end{equation}
The interpolation between
$a'_{\mathrm{BL,I}} \to a'_{\mathrm{BL,II}}$
[see Eqs.~\eqref{coeffBLI} and~\eqref{coeffBLII}],
consistent with Eq.~\eqref{aprimeBL},
can be traced to the algebraic identity
\begin{equation}
\frac{11}{336} \times \left( 1 + 
\frac{4}{11} \right) = \frac{5}{112}
\qquad
\mbox{[$a'_{\mathrm{BL,I}} \to a'_{\mathrm{BL,II}}$] \,,}
\end{equation}
which we would also like to indicate, for 
full clarification.}

Note that the coefficients proportional to $\del_e (\del_i - \del_e)$
vanish in the model proposed by Bezrukov and Lee~\cite{BeLe2012}.
Turning to the case of ``slow'' superluminal neutrino pair creation by 
emission from a ``fast'' superluminal neutrino, i.e., $\psi_f = \nu_f = \nu^{(m)}_f$, 
we again parameterize the total decay width and differential energy loss as
\begin{align}
\label{NPCRfunc}
\Gamma_{\nu_i \to \nu_i \nu_f \bar{\nu}_f} 
&= b \frac{G_F^2}{192\pi^3} k_1^5\,,
\\
\frac{\dd E_{\nu_i \to \nu_i \nu_f \bar{\nu}_f}}{\dd x} &= 
-b' \frac{G_F^2}{192\pi^3} k_1^6\,.
\end{align}
In this case too, we can consider the prescription 
of Ref.~\cite{CoGl2011} for the spin sums
and the interaction Lagrangian with $v_{\rm int} = 0$. Then we find
\begin{subequations}
\label{resbCG}
\begin{align}
b_{\mathrm{CG}} =& \; \frac{1}{7} (\del_i - \del_f) 
\left[(\del_i - \del_f)^2 + \frac{5}{3} \del_f (\del_i - \del_f) 
+ \frac{7}{15}\del_f^2\right]\,,
\\
b'_{\mathrm{CG}} =& \; \frac{25}{224} (\del_i - \del_f) 
\left[(\del_i - \del_f)^2 \right.
\nonumber\\[0.1133ex]
& \; \left. 
+ \frac{112}{75} \del_f (\del_i - \del_f) + \frac{32}{75}\del_f^2\right]\,.
\end{align}
\end{subequations}
These results are applicable for $\delta_i > \delta_f$, as explained above.
(Otherwise, the available phase space for the outgoing particles vanishes.)
We have checked that the $\del_f \to 0$ limit corresponds 
to the previous result for $\psi_f=e$ after 
accounting for the different factors of $f_e$, $c_A$, $c_V$ as well as 
the number of spin states $n_s$, 
\color{black} 
which we take as $n_s = 2$ for the 
model of Cohen and Glashow, and as $n_s=1$ for the models of Bezrukov and Lee,
in accordance with the prescriptions used
in Refs.~\cite{CoGl2011,BeLe2012}.
\color{black}

We can also insert the spin sum prescription from 
Bezrukov and Lee, \Refr{BeLe2012}, Eq.~\eqref{BLspinsum},
and use the metric ${\tilde g}_{\lambda\sigma}(v_{\rm int})$ given in 
Eq.~\eqref{tildeg} with $\delta_{\rm int} = v_{\rm int}^2 - 1$ 
left as a free parameter, in the interaction 
Lagrangian~\eqref{Lint}.
In this case we obtain
\begin{subequations}
\label{resbBL}
\begin{align}
b_{\mathrm{BL}} =& \; \frac{17}{210} (\del_i - \del_f) 
\left[(\del_i - \del_f)^2 + \frac{7}{17}\del_{\rm int}^2\right]\,,
\\
b'_{\mathrm{BL}} =& \; \frac{11}{168} (\del_i - \del_f)
\left[(\del_i - \del_f)^2 + \frac{4}{11}\del_{\rm int}^2\right]\,.
\end{align}
\end{subequations}
Better (but not full) compliance with $SU(2)_L$ gauge invariance is ensured
by setting $\delta_{\rm int} = \delta_i + \delta_f$
(see Appendices~\ref{appa} and~\ref{appb}).

Finally, we offer a comment on the LPCR process assuming that also the 
electron is slightly superluminal. Assuming that $\delta_e$ is of the same 
order of magnitude as $\delta_\nu$ and $\delta_{\rm int}$, Eqs.~\eqref{eq:Gamma-general} 
and \eqref{eq:dEdx-general} demonstrate that the 
results presented above in Eqs.~\eqref{resbCG} 
and \eqref{resbBL} (valid to leading order in 
the small $\delta$'s) for the NPCR process are 
only modified with respect to LPCR due to the different 
values of $f_e$ [see Eq.~\eqref{eq:fe}] and the couplings 
$c_A$ and $c_V$ [see Eq.~\eqref{eq:cAcV}]. Accounting for these differences 
simply amounts to dividing the results of Eqs.~\eqref{resbCG} and \eqref{resbBL} 
by a factor of 
two. We note that after taking this extra factor of $1/2$ into account, 
the $\del_f \to 0$ limit of Eqs.~\eqref{resbCG} and \eqref{resbBL} indeed 
reproduces the results of Eq.~\eqref{resa} with $\delta_{\rm int}$ chosen 
appropriately: $\delta_{\rm int} = 0$ for the Cohen-Glashow model,
and model I of 
Bezrukov and Lee, while $\delta_{\rm int} = 
\delta_i + \delta_f \to \delta_i$ for model II of 
Ref.~\cite{BeLe2012}.

%
%
\subsection{Interpretation of the Results}
\label{sec42}

Let us try to interpret the results presented 
in Sec.~\ref{sec41}, both qualitatively and quantitatively.

The first point to address is the fractional energy
loss during an NPCR decay process.
The double-differential
energy loss with respect to energy and time,
due to VPE or neutrino splitting, is
\begin{equation}
\dd E_1 = (E_1 - E_3) \frac{\dd \Gamma}{\dd E_3} 
\dd t = (E_1 - E_3) \frac{\dd \Gamma}{\dd E_3} \frac{\dd x}{c} \,,
\end{equation}
where we restore the factor of $c$ for clarity.
One then obtains, in natural units,
\begin{equation}
\frac{\dd E_1}{\dd x} = \int_0^{E_1} (E_1 - E_3) 
\frac{\dd \Gamma}{\dd E_3}\, \dd E_3 \,,
\end{equation}
but the expression of the right-hand side of this
formula has the alternative interpretation as the 
mean energy loss during a decay event. Hence,
the average fractional energy loss during a decay 
event can be evaluated as
\begin{equation}
f = \frac{1}{E_1 \, \Gamma} \, \frac{\dd E_1}{\dd x} \,,
\end{equation}
which is a constant for all processes 
described by Eqs.~\eqref{resLPCR}---\eqref{resbBL}.
For LPCR, in the Cohen--Glashow formulation~\cite{CoGl2011},
one obtains $f = (25/224)/(1/7) = 25/32 = 0.78$,
which explains the remark made in Refs.~\cite{StSc2014,StEtAl2015}
regarding an average 78\,\% energy loss during 
each single vacuum pair emission event.

For the term proportional to $(\delta_i - \delta_f)^3$
in NPCR, still within the Cohen--Glashow formulation~\cite{CoGl2011},
one obtains the same ratio 
$f = (25/112)/(2/7) = 0.78$,
while for the term proportional to $(\delta_i - \delta_f) \, \delta_f^2$,
the result is $f = 7/10 = 0.7$.
For the model II of Bezrukov and Lee~\cite{BeLe2012},
one obtains ratios of $f = 55/68 = 0.81$
[for the $(\delta_i - \delta_f)^3$ term]
and $f = 5/7 = 0.71$
[for the $(\delta_i - \delta_f) \, \delta_{\rm int}^2$ term].
At variance with these results, in Ref.~\cite{StEtAl2015},
an equipartition of energy among the
products of the neutrino splitting is assumed,
based on a Planck-scale generated, dimension-six
operator. 

Let us investigate if the NPCR process could 
still generate a substantial contribution to 
neutrino energy losses on astrophysical time scales,
given the tight constraints on the difference between 
Lorentz-violating parameters coming from
short-baseline~\cite{AAEtAl2013boone} and
``extremely long-baseline'' 
experiments (Ref.~\cite{AaEtAl2010lorentz}).

The model preferred in Ref.~\cite{StEtAl2015} 
concerns a dimension-six operator which leads to 
a decay rate that has the same functional dependence 
as Eq.~\eqref{NPCRfunc}, 
but with the $b$ parameter replaced by
an expression proportional to 
\begin{equation}
b \propto \left( \frac{k_1}{M_{\rm Pl}} \right)^6 \,,
\end{equation}
corresponding to a replacement of the 
term $(\delta_i - \delta_f) \, \delta_f^2 $ by an expression
proportional to 
\begin{equation}
(\delta_i - \delta_f) \, \delta_f^2 
\propto \left( \frac{k_1}{M_{\rm Pl}} \right)^6 \,.
\end{equation}
(This formula should be compared to a potential
energy dependence of Lorentz-violating parameters,
as envisaged in Ref.~\cite{AlElMa2012}.)

The surprising conclusion is as follows.
Even at the fantastically small differential
Lorentz violations of $\delta_i - \delta_f < 7.4 \times 10^{-27}$
(see Ref.~\cite{AaEtAl2010lorentz}),
the NPCR process can still substantially contribute to the
energy loss of neutrinos on astrophysical scales,
if we use $\delta_f \sim 1.0 \times 10^{-20}$ 
(Refs.~\cite{StSc2014,StEtAl2015}).
Namely, at an energy of $10 \,{\rm PeV}$, which 
according to Refs.~\cite{StSc2014,StEtAl2015}
is commensurate
with a cutoff of the astrophysical neutrino spectrum
at about $2 \,{\rm PeV}$ (arriving on Earth),
our process induces a decay rate, and an energy loss rate,
commensurate with
a Planck-scale, dimension-six operator with a large numerical coefficient
$\kappa_2 = 258$ (in the conventions of Ref.~\cite{StEtAl2015}).
For $\delta_i - \delta_f \lesssim 6 \times 10^{-22}$
(see Ref.~\cite{CoGl1999}),
a Planck-scale, dimension-six operator with a large numerical coefficient
of $\kappa_2 = 9200$ would be required to 
lead to a comparable effect at 
the quoted energy of $10 \,{\rm PeV}$.
For comparison, we note that 
in order to explain a putative cutoff 
of cosmic neutrinos at energy $2 \, {\rm PeV}$,
a value of $\kappa_2 = 7800$ is otherwise
required~\cite{StEtAl2015}.
This observation illustrates that
the final evaluation of the relevance
of the NPCR process could depend on the
clarification of the precise location,
and the physical mechanism behind the
conjectured cutoff of the high-energy
cosmic neutrino spectrum.

A semi-quantitative observation is of interest.
Namely, a signature of the NPCR process
would be departure from a $(1:1:1)_\oplus$ 
equipartition of neutrino flavors arriving on Earth,
due to a decay of all neutrino
mass eigenstates except the slowest one,
above a certain energy scale where the
decay channel becomes numerically relevant.
(This is perhaps not so evident for the 
dimension-six operator verified in Ref.~\cite{StEtAl2015},
but evident for our scenario studied here.
The general idea that neutrino decays could 
alter the flavor composition arriving on Earth
has been formulated in Ref.~\cite{BeEtAl2003}.)
An analysis of neutrinos above
$35\, {\rm TeV}$ arriving at IceCube~\cite{AaEtAl2013flavor}
(see also Refs.~\cite{BuBeWi2015,ArKaSa2015,LiBuBe2016}) 
is statistically compatible with an equal flavor
distribution $(1:1:1)_{\oplus}$ on Earth,
but in the caption of Fig.~3 of Ref.~\cite{AaEtAl2013flavor},
it is explicitly stated that the
best-fit composition at Earth is
$(0 : 1/5 : 4/5)_\oplus$.
In the caption of Fig.~5 of the recent
work~\cite{AaEtAl2018},
the best-fit composition is given as
$(0 : 0.21 : 0.79)_\oplus$.
Let us have a look at the
structure of the PMNS matrix, which is given as follows,
\begin{equation}
U = \left(
\begin{array}{ccc}
U_{e1} & U_{e2} & U_{e3} \\
U_{\mu 1} & U_{\mu 2} & U_{\mu 3} \\
U_{\tau 1} & U_{\tau 2} & U_{\tau 3} \\
\end{array}
\right) \,.
\end{equation}
Modulus-wise, one has very large mixing
[see Eq.~(2.2) of Ref.~\cite{EsEtAl2018}],
\begin{multline}
|U| = \left(
\begin{array}{ccc}
|U_{e1}| & |U_{e2}| & |U_{e3}| \\
|U_{\mu 1}| & |U_{\mu 2}| & |U_{\mu 3}| \\
|U_{\tau 1}| & |U_{\tau 2}| & |U_{\tau 3}| \\
\end{array}
\right)
\\
= \left(
\begin{array}{ccc}
0.797\ldots 0.842 &  0.518\ldots 0.585 &  0.143\ldots 0.156 \\
0.235\ldots 0.484 &  0.458\ldots 0.671 &  0.647\ldots 0.781 \\
0.304\ldots 0.531 &  0.497\ldots 0.699 &  0.607\ldots 0.747 \\
\end{array}
\right) .
\end{multline}
The ``second'' mass eigenstate thus has
the roughly equal flavor decomposition
\begin{equation}
\nu^{(m)}_2 =
U^*_{e2} \, \nu^{(f)}_e +
U^*_{\mu 2} \, \nu^{(f)}_\mu +
U^*_{\tau 2} \, \nu^{(f)}_\tau \,,
\end{equation}
where
$| U^*_{e2} | \approx
| U^*_{\mu 2} | \approx
| U^*_{\tau 2} | \approx
1/\sqrt{3}$
(within numerical uncertainty),
while mass eigenstate ``number 3'' leans more
toward a higher $\mu$-neutrino and $\tau$-neutrino
content, consistent with the trend
of the data reported in Fig.~3 of
Ref.~\cite{AaEtAl2013flavor}.
In principle, one could thus speculate
about mass eigenstate ``number 3'' serving as
the ``slowest'' mass eigenstate,
which thus would be the only one not
affected by neutrino splitting.
However, at present, this observation
does not go beyond the status of pure speculation,
in the sense of ``reading the tea leaves''.
In particular, the parameters required for the NPCR
process, as formulated here, are constrained
to a kinematic region incompatible with
a conjectured caused cutoff of all neutrino
mass eigenstates except the slowest one,
at a comparatively low energy of only $35 \, {\rm TeV}$.

\begin{figure} [t]
\begin{center}
\begin{minipage}{0.99\linewidth}
\begin{center}
\includegraphics[width=0.91\linewidth]{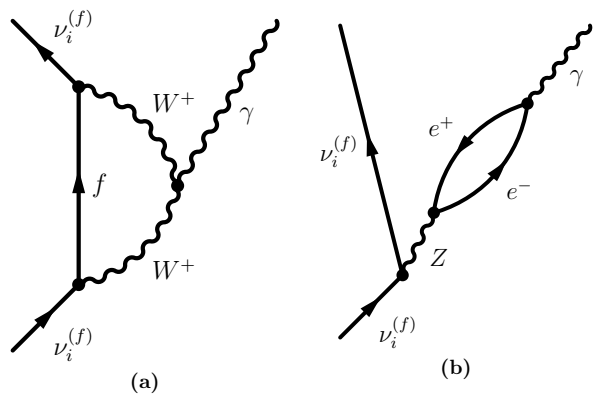}
\caption{\label{fig2}
In the genuine Cerenkov radiation process 
$\nu \to \nu + \gamma$ (for superluminal neutrinos), the 
photon is emitted from a $W$ loop 
or from a vacuum-polarization insertion 
into the $Z$ propagator;
the emission takes place from a flavor 
eigenstate $| \nu^{(f)}_i \rangle$.
The photon emission thus involves an 
extra factor $\alpha_{\rm QED} \approx 
1/137.036$ at the $W$--photon vertex
as compared to LPCR and NPCR.}
\end{center}
\end{minipage}
\end{center}
\end{figure}

Likewise, it is instructive to compare 
the consequences of the neutrino splitting 
and its signatures to those discussed in Ref.~\cite{MoSa2018},
where the authors
present calculations for the flavor ratio of IceCube in the 
high-energy region, given the putative presence of a 
$\nu \to 3 \nu$ neutrino splitting mechanism
in the high-energy region, albeit a different 
one as compared to the one discussed here, namely,
mediated by a second Higgs doublet (box diagrams).
The authors of Ref.~\cite{MoSa2018} consider a model based on the symmetry
group $SU(3)_c \times SU(2)_L \times U(1)_Y \times Z_2$,
which implies that for any 
usual Standard Model fermion, 
one has one electroweak singlet right-handed neutrino.
The model is referred to as the $\nu$2HDM 
(neutrino--two--Higgs doublet model).

In the text following Eq.~(4.3) of Ref.~\cite{MoSa2018},
the authors present predictions for 
neutrino flavor ratios, based on additional
assumptions on the production mechanism and the 
mass hierarchy. 
They ascertain that, for normal mass hierarchy,
neutrinos coming from pion decay split,
according to $\nu \to 3\nu$, 
to produce flavor ratios of daughter 
neutrinos in the flavor basis as
$\phi_e : \phi_\mu : \phi_\tau \approx 
2 : 1 : 1$.
For neutrinos coming from muon damped sources,
still with normal mass hierarchy, the prediction is 
$\phi_e : \phi_\mu : \phi_\tau \approx 
1.8 : 1 : 1$.
For neutrinos from neutron decay, the
splitting is
$\phi_e : \phi_\mu : \phi_\tau \approx 
3 : 1 : 1$.

None of the predictions presented in Ref.~\cite{MoSa2018}
is consistent with a flavor ratio that would corresponds to 
exactly one of the mass eigenstates of the PMNS matrix;
the latter would otherwise be predicted
by the Lorentz-violation mediated
neutrino splitting model.
Namely, as already stressed,
the superluminal model 
would predict that only one mass eigenstate survives.
So, even though questions regarding the 
mass hierarchy of the neutrino mass eigenstates
have not been conclusively addressed,
we can say that the model discussed in 
our work, and the one discussed in Ref.~\cite{MoSa2018},
have distinctly different signatures.

%
%
\section{Conclusions}
\label{sec5}

Let us summarize the most important findings reported here.

{\em (i)} Three decay processes have been identified in 
Ref.~\cite{CoGl2011} which become kinematically 
possible for an incoming, superluminal, Lorentz-violating
neutrino: LPCR [Eq.~\eqref{LPCR}], NPCR [Eq.~\eqref{NPCR}], 
as well as the process $\nu \to \nu + \gamma$;
the latter constitutes genuine Cerenkov radiation in vacuum
(see Fig.~\ref{fig2}).
For $\nu \to \nu + \gamma$, the photon is emitted from a $W$ loop
or from a vacuum-polarization correction to the $Z$ propagator,
and thus involves an extra factor $\alpha_{\rm QED} \approx 1/137.036$.
This process is thus parametrically suppressed.
Here, we show that the NPCR process is not
parametrically suppressed in comparison to LPCR.
Furthermore, the threshold for NPCR is at least six orders 
of magnitude lower than for LPCR [Eq.~\eqref{EthNPCR}].
We assume that 
the velocity parameters $v_i$ in the dispersion relation~\eqref{disp_rel}
for the three mass eigenstates are not all identical.
While the relative differences among Lorentz-violating
parameters for the neutrino mass and flavor
eigenstates are otherwise tightly
constrained~\cite{CoGl1999,AaEtAl2010lorentz,%
AAEtAl2013boone},
our assumption is supported by the fact that the corresponding
mass terms $m_i$ in Eq.~\eqref{disp_rel} 
also are different from each other.

{\em (ii)} We here confirm the results obtained for 
LPCR obtained in Refs.~\cite{CoGl2011,BeLe2012},
including the model dependence derived in Ref.~\cite{BeLe2012}.
This also reaffirms the validity of the astrophysical 
bounds on the Lorentz-violating parameters, 
derived in Refs.~\cite{StSc2014,StEtAl2015},
based on dimension-four operators.
Our expressions for the NPCR process
are parametrically of the same order as for LPCR,
but the overall coefficients are larger by a factor 
four or five. We should also point out the somewhat unexpected 
terms in the NPCR decay rates reported in Eqs.~\eqref{resbCG},
and~\eqref{resbBL}, for Lorentz-violating neutrinos,
proportional to $(\delta_i - \delta_f) \, \delta_f^2$
instead of $(\delta_i - \delta_f)^3$.
We also derive a few more pieces of 
information, e.g., the departure from the equipartition 
of energy between the decay products of neutrino splitting,
with the incoming neutrino being shown to lose about (75$\pm$5)\%
of its energy during NPCR decay.
The structure of the dimension-four NPCR operators
derived here also raises a pertinent question regarding the 
possible presence of terms proportional to 
$\Theta(\delta_i - \delta_f)$ (where $\Theta$ is the 
Heaviside step function) in the dimension-five
and dimension-six operators, which were derived
from Planck-scale physics (see Ref.~\cite{StEtAl2015}).

{\em (iii)} If, in the future, hypothetically,
the high-energy behavior of neutrinos
should be confirmed to be superluminal and Lorentz violating,
but with exceedingly small parameters,
then our results will help in the modeling of
the influence of the LPCR process on intergalactic
neutrino propagation.

Any statement beyond the above observations
would require an elaborate
Monte Carlo simulation of astrophysical data
(see Refs.~\cite{StSc2014,StEtAl2015}),
which is beyond the scope of the current paper.
A general picture emerges from the analysis of 
LPCR and NPCR for high-energy incoming, superluminal 
neutrinos: namely, at high energy, even very tiny 
parameters $\delta$ lead to a high virtuality
$E^2 - \vec p^{\,2} = 
(v^2 - 1) \vec p^{\,2} \approx
\delta \times E^2$, which grows with the energy.
A high virtuality implies that various pair 
production processes become kinematically 
possible in the high-energy domain.
This observation leads to the very strict 
bounds on the Lorentz-violating parameters 
and puts very tight 
constraints on the Lorentz-violating models.
Indeed, the LPCR and NPCR process, as well as vacuum Cerenkov
radiation, imply very tight restrictions on the available 
parameter space for Lorentz violation in
the high-energy neutrino sector.

\section*{Acknowledgements}

The authors would like to mention insight
discussions with, and advice from, A.~de Gouvea.
Support from the National Science Foundation (Grant
PHY--1710856) is gratefully acknowledged.
This work was also supported by the \'UNKP-17-3 New National Excellence
Program of the Ministry of Human Capacities of Hungary
and by grant K 125105 of the National Research, Development
and Innovation Fund in Hungary.

\appendix

%
%
\section{Generalized Dirac Equations}
\label{appa}

It is instructive to consider the derivation 
of the Lorentz-violating dispersion relation~\eqref{disp_rel}
on the basis of a generalized Dirac equation,
as well as the statement made after Eq.~\eqref{Lint},
which concerns the fact that the 
Lagrangian in Eq.~\eqref{Lint}, for $v_{\rm int} = v_i$,
can be considered as an $SU(2)_L$-gauge invariant 
Lagrangian, whereas for $v_{\rm int} = 1$ ($v_{\rm int} = c$),
or $\delta_{\rm int} = 0$, it is not gauge invariant.
In order to proceed with the proof,
let us denote by 
\begin{equation}
g_{\mu \nu} = {\rm diag}(1, -1, -1, -1)
\end{equation}
the flat-space, standard space-time metric, and by 
${\tilde g}_{\mu \nu}$ a generalized metric 
of constant coefficients, which parameterizes the 
Lorentz violation. 
It {\em can} take the form [see Eq.~\eqref{tildeg}]
\begin{equation}
\label{example}
{\tilde g}_{\mu \nu}(v^2) = {\rm diag}(1, -v^2, -v^2, -v^2) \,,
\end{equation}
where we note the square of the velocity,
but the formalism outlined below is more 
general. We define the generalized Dirac matrices
[cf.~Refs.~\cite{Je2013,JeNo2013pra,JeNo2014jpa,NoJe2015tach,NoJe2016,Je2018geonium}]
\begin{equation}
\label{def_vierbein}
{\tilde \gamma}_\mu = e_\mu^A \, \gamma_A \,,
\end{equation}
where the Einstein summation convention is 
used, and $\gamma^A$ with $A = 0,1,2,3$ 
are the ordinary Dirac $\gamma$ matrices, while
the $e_\mu^A$ take the role of the 
so-called ``vierbein'' in general relativity,
with the property
\begin{equation}
{\tilde g}_{\mu\nu} 
= e_\mu^A \, g_{AB} \, e_\nu^B 
= e_\mu^A \, e_{\nu A} \,.
\end{equation}
This implies that 
the ``vierbein'' takes the role of the square root of the 
metric~\cite{JeNo2013pra}].
Capital Latin indices can be raised with the flat-space metric $g^{AB}$.
One can then easily show that
\begin{equation}
\{ {\tilde \gamma}_\mu, {\tilde \gamma}_\nu \} =
e_\mu^A \, e_\nu^B \, \{ \gamma_A, \gamma_B \} =
e_\mu^A \, e_\nu^B \, ( 2 g_{A B} ) =
2 {\tilde g}_{\mu\nu} \,.
\end{equation}

The analogy to the formalism
of general relativity implies that 
${\tilde g}_{\mu \nu}$ takes the role 
of a modified Lorentz ``metric'', but
without curvature (because we assume that the 
coefficients are constant).
The word ``metric'' should be understood 
with a grain of salt (hence the apostrophes),
because it does not constitute a space-time metric 
in the sense of general relativity,
that is used to measure space-time intervals,
but rather, a mathematical object used to 
parameterize the dispersion relation 
of a Lorentz-violating particle.
Because of the lack of 
curvature, the ``metric'' ${\tilde g}_{\mu \nu}$ 
is still characterizing a flat ``space-time''.
(For a truly curved space, the 
notation ${\overline g}_{\mu\nu}$ has been 
proposed in Refs.~\cite{Je2013,JeNo2013pra} in order
to distinguish the curved-space quantities
from the flat-space ones.)
The $\tilde\gamma_\mu$ are thus intermediate in between the
usual flat-space Dirac matrices, and the curved-space
matrices, for which the notation $\overline g_{\mu\nu}$
has been proposed in 
Refs.~\cite{Je2013,JeNo2013pra,JeNo2014jpa,NoJe2015tach,NoJe2016,Je2018geonium}.

For a modified ``metric'' of the form~\eqref{example},
one can choose the vierbein coefficients as
\begin{equation}
\label{vier_value}
e^0_0 = 1 \,,
\quad
e^0_i = e^i_0 = 0 \,,
\quad
e^i_j = v \, \delta^i_j \,,
\quad 
i,j = 1,2,3 \,.
\end{equation}
The modified Dirac equation describing the Lorentz
violation can then be written as
\begin{equation}
\left( \ii \, {\tilde \gamma}_\mu \, 
\partial^\mu - m \right) \, \psi = 0 \,.
\end{equation}
We here suppress the chirality projectors
and assume that $\psi$ is a left-handed field.
One can multiply from the left by the operator
$[\ii \, {\tilde \gamma}_\nu \, \partial^\nu + m]$,
and use the operator identity
\begin{equation}
\left( \ii \, {\tilde \gamma}_\nu \,
\partial^\nu + m \right) \, 
\left( \ii \, {\tilde \gamma}_\mu \,
\partial^\mu - m \right) 
= -{\tilde g}_{\mu \nu} \,
\partial^\mu \, \partial^\nu - m^2 \,.
\end{equation}
For the metric~\eqref{example}, one can use the 
identity
\begin{equation}
-{\tilde g}_{\mu \nu} \,
\partial^\mu \, \partial^\nu - m^2 = 
E^2 - v^2 \, \vec p^{\,2} - m^2 \,,
\end{equation}
where $E$ is the energy and $\vec p$ is the momentum
operator. This leads to the dispersion relation,
\begin{equation}
E = \pm \sqrt{\vec p^{\,2} \, v^2 + m^2} \,,
\end{equation}
which is equivalent to Eq.~\eqref{disp_rel}.
\color{black}
We here note, though, that Eq.~\eqref{disp_rel} 
has no gauge structure; it is formulated for 
free particles.
\color{black}

%
%
\section{Gauge Invariance of the Models}
\label{appb}

\color{black}

For a full clarification of the models
used in the current investigation, 
and in order to avoid misunderstandings,
a number of remarks on the 
gauge (non-)invariance of the interaction 
Lagrangian~\eqref{Lint} are in order.
Of course, the interaction
Lagrangian~\eqref{Lint} in 
itself of course does not describe
a gauge-invariant theory, but merely 
constitutes the low-energy limit of the 
full electroweak theory. It is 
still applicable for the description 
of the decay and energy loss processes analyzed in the current paper
[see also the remarks in the text 
preceding Eq.~\eqref{Lint}].

Furthermore, the velocity parameter 
$v_{\rm int}$ that enters Eq.~\eqref{Lint}
of course depends on the details of 
the gauge-invariant theory that one started 
from, in the derivation of the 
effective ``low-''energy interaction
given in Eq.~\eqref{Lint}.

We should mention that 
questions related to the gauge (non-)invariance 
of the models have recently been 
analyzed in detail, in a separate paper~\cite{JeNaSo2019}.
We would like to provide a summary here.
In the model used by Cohen and Glashow~\cite{CoGl2011},
which corresponds to model~I used by Bezrukov
and Lee~\cite{BeLe2012}, 
and also corresponds to the interaction Lagrangian~\eqref{Lint}
with $v_{\rm int} = 1$, it cannot be overemphasized that 
$SU(2)_L$ gauge invariance is manifestly 
broken. In fact, model I used by Bezrukov
and Lee~\cite{BeLe2012} is obtained if one 
postulates that the Lorentz-breaking 
term introduced into the free Dirac equation for the 
Lorentz-violating neutrinos
is not ``gauged'', i.e., retains the partial
derivative as opposed to the 
covariant derivative in the neutrino sector. 
In this case, the interaction Lagrangian
is unaltered in comparison to the Fermi theory,
while the free Lagrangian of the neutrinos acquires 
a Lorentz-violating term.
In order to put things into perspective,
one should note that the gauge dependence in model I 
enters only at the
perturbative level, i.e., on the same level as the
Lorentz-violating operator itself enters the Lagragian~\cite{JeNaSo2019}.
In order to verify that model~I is based on reasonable 
assumptions, one can point to the observations made in
papers of Nielsen {\em et al.}~\cite{ChFrNi2001prl,ChFrNi2001npb,Bj2001,AzCh2006,%
ChJe2007,ChFrJeNi2008,ChJe2008,ChFrNi2009},
where the authors (in a somewhat different context) 
observe the emergence of 
gauge-symmetry breaking terms, upon the introduction
of (initially) spontaneous Lorentz-(but not gauge)-symmetry
breaking. In this context,
we note that the three-photon vertex, and the
two-fermion, two-photon interaction
in the Lagrangian given in Eq.~(3) of Ref.~\cite{ChJe2008},
which are initially generated by spontaneous
Lorentz breaking in the electromagnetic sector, 
break electromagnetic gauge invariance.
With reference to Eqs.~(9) and (10) 
of Ref.~\cite{CoKo1998}, it cannot be stressed enough
that the model of used by Cohen and Glashow 
in Ref.~\cite{CoGl2011} is {\em outside} 
of the original formulation of the Standard Model Extension~\cite{CoKo1998}.

Even model~II used by Bezrukov and Lee~\cite{BeLe2012},
somewhat interestingly, is not gauge invariant
with respect to the full $SU(2)_L \times U(1)_Y$ gauge theory
(see Sec.~IV of Ref.~\cite{JeNaSo2019}),
but follows from the full gauge theory if
one reduces the gauge group to $U(1) \times U(1)_Y$,
i.e., to a sector where only the interaction 
terms corresponding to the $Z$ boson and the the
photon are (re-)diagonalized (but not the $W$ boson
interaction terms).
The partial retention of gauge invariance
justifies, {\em a posteriori}, to a certain degree,
the statement made by Bezrukov and Lee that their
model II is ``gauge invariant'' where we note 
the quotation marks in the text after Eq.~(4) of Ref.~\cite{BeLe2012},
implying that their statement should be taken {\em cum grano salis}.
Again,  with reference to Eqs.~(9) and (10)
of Ref.~\cite{CoKo1998}, model II used by 
Bezrukov and Lee~\cite{BeLe2012} also is {\em outside}
of the original formulation of the Standard Model Extension,
but implements a restricted set of symmetry groups:
Namely, it reduces the Lorentz symmetry group 
from $SO(1,3)$ to $SO(3)$, in view of the spatially 
isotropic Lorentz violation, and the 
electroweak gauge group from $SU(2)_L \times U(1)_Y$ 
to $U(1) \times U(1)_Y$. Model~II thus constrains 
Lorentz-breaking and gauge-symmetry breaking parameters 
within the given restricted symmetry groups.

As a last point, we mention that,
as pointed out in Sec.~V of Ref.~\cite{JeNaSo2019},
there might actually be a possibility to formulate a
Lorentz-violating theory, which fully preserves
$SU(2)_L \times U(1)_Y$ gauge invariance and still
allows for the LPCR and NPCR decays. The decisive
idea is to postulate Lorentz-violating parameters
which depend on the flavor (see Sec.~V of Ref.~\cite{JeNaSo2019}).
We recall that,
if all left-handed charged fermions and all neutrinos
are grouped together in $SU(2)_L$ multiplets and if
there is uniform Lorentz violation over all
generations (flavors), then both NPCR as well
as LPCR are kinematically forbidden.
However, if the Lorentz-violating parameters
are different among the fermion flavors, then
the processes become kinematically allowed,
and under the given assumptions, with a fully $SU(2)_L
\times U(1)_Y$ covariant coupling,
the form of the interaction Lagrangians
is uniquely determined. Details on further calculations
based on these models,
including additional proofs regarding the 
gauge invariance with respect to the 
particular choice for the gauge boson propagator,
will be published elsewhere~\cite{SoJeNa2019prep}.

\color{black}

\end{document}